\documentclass[12pt]{iopart}
\usepackage{iopams}  
\usepackage{bm}
\usepackage{graphicx}

\begin{document}

\title[Relativistic probability densities for location]{
Relativistic probability densities for location}

\author{Joshua G.~Fenwick, Rainer Dick}

\address{Department of Physics and Engineering Physics, 
University of Saskatchewan, 116 Science Place, Saskatoon, Canada SK S7N 5E2}

\begin{abstract} 
Imposing the Born rule as a fundamental principle of quantum mechanics would require
the existence of normalizable wave functions $\psi(\bm{x},t)$ also for relativistic particles.
Indeed, the Fourier transforms of normalized $\bm{k}$-space
amplitudes $\psi(\bm{k},t)=\psi(\bm{k})\exp(-\,\mathrm{i}\omega_{\bm{k}}t)$ yield
normalized functions $\psi(\bm{x},t)$ which reproduce the standard $\bm{k}$-space
expectation values for energy and momentum from local momentum (pseudo-)densities 
$\wp_\mu(\bm{x},t)=(\hbar/2\mathrm{i})[\psi^+(\bm{x},t)\partial_\mu\psi(\bm{x},t)
  -\partial_\mu\psi^+(\bm{x},t)\cdot\psi(\bm{x},t)]$. However, in the case of
bosonic fields,
the wave packets $\psi(\bm{x},t)$ are nonlocally related to the corresponding
relativistic quantum fields $\phi(\bm{x},t)$,
and therefore the canonical local energy-momentum
densities $\mathcal{H}(\bm{x},t)=c\mathcal{P}^0(\bm{x},t)$
and $\bm{\mathcal{P}}(\bm{x},t)$ differ from $\wp_\mu(\bm{x},t)$ and
appear nonlocal in terms of the wave packets $\psi(\bm{x},t)$. 
We examine the relation between the canonical energy density
$\mathcal{H}(\bm{x},t)$, the canonical charge density $\varrho(\bm{x},t)$, the energy
pseudo-density $\tilde{\mathcal{H}}(\bm{x},t)=c\wp^0(\bm{x},t)$,
and the Born density $|\psi(\bm{x},t)|^2$ for the massless free Klein--Gordon field.
We find that those four proxies for particle location are tantalizingly close
even in this extremely relativistic case:
In spite of their nonlocal mathematical relations,
they are mutually local
in the sense that their maxima do not deviate beyond a common position uncertainty $\Delta x$.
Indeed, they are practically indistinguishable in cases where we would expect a normalized
quantum state to produce particle-like position signals, \textit{viz.}~if
we are observing quanta with momenta $p\gg\Delta p\ge\hbar/2\Delta x$.

We also translate our results to massless Dirac fields.
Our results confirm and illustrate that the normalized energy
density $\mathcal{H}(\bm{x},t)/E$ provides a suitable measure for positions of bosons,
whereas normalized charge density $\varrho(\bm{x},t)/q$ provides a suitable measure for
 fermions.
\end{abstract}


%
\vspace{2pc}

\section{Introduction\label{sec:intro}}

Both relativistic bosons and fermions can generate well-localized tracks
in particle detectors. For example, ultrarelativistic electrons with
energies $E<3m_ec^2$ are stable
against decays $e^-\to e^-e^-e^+$, and they are capable of producing long particle
tracks as long as there are no positrons around. They are also inert to
pair-creation during scattering, $e^-e^-\to e^-e^-e^-e^+$, when
scattering off nonrelativistic electrons in a detector.

The same remarks apply to $\pi^+$ mesons with energies $E<3m_{\pi} c^2$, which can
generate 10-metre tracks before decaying if their $\gamma$ factor
is $\gamma\lesssim 3$. Furthermore, position resolutions for
photon absorption in superconducting nanowire single-photon detectors have 
been reported with a spatial resolution
of $12.6\times 12.6\,\mathrm{micrometre}^2$ \cite{naturephotonics},
which is compatible with the best position resolutions in charged particle
detectors \cite{pdg}.

Therefore, although relativistic quantum theory is inherently a many-particle
theory, the fact that quantum field theory includes particle-antiparticle
reactions and particle decays does
not excuse us from identifying a measure for the probability that a relativistic
charged particle or a photon create a signal ``here'' but not ``there''.

Surprisingly, but likely also inevitably, the discussion of relativistic probability
measures for particle location has led to different results for bosons
and for fermions.

To explain this for fermions, we write the Dirac field 
\begin{eqnarray}\nonumber
\varphi(\bm{x},t)&=&\frac{1}{\sqrt{2\pi}^3}\int\!
d^3\bm{k}\sum_{s}
\bigg(\psi_s(\bm{k})u(\bm{k},s)\exp[\mathrm{i}(\bm{k}\cdot\bm{x}-\omega_{\bm{k}}t)]
\\ \label{eq:diracfs1}
&&+\chi_s^+(\bm{k})v(\bm{k},s)\exp[-\,\mathrm{i}(\bm{k}\cdot\bm{x}-\omega_{\bm{k}}t)]
\bigg)
\end{eqnarray}
with normalized 4-spinors (see e.g.~Eqs.~(\ref{eq:u1ks}-\ref{eq:v1ks}) below)
\begin{equation}
u^+(\bm{k},s)\cdot u(\bm{k},s')=\delta_{ss'},\quad
v^+(\bm{k},s)\cdot v(\bm{k},s')=\delta_{ss'},
  \end{equation}
\begin{equation}
u^+(\bm{k},s)\cdot v(-\,\bm{k},s')=0.
  \end{equation}
A single-particle amplitude for the Dirac field would be given by
\begin{equation}\label{eq:1pdirac}
\int\!d^3\bm{k}\sum_s|\psi_s(\bm{k})|^2=1,\quad\chi_s(\bm{k})=0.
\end{equation}
This field would carry energy 
\begin{eqnarray} \label{eq:Edirac}
E&=&\int\!d^3\bm{x}\,\mathcal{H}_{\varphi}(\bm{x},t)
=\int\!d^3\bm{k}\,\hbar\omega_{\bm{k}}\sum_s|\psi_s(\bm{k})|^2
  \end{eqnarray}
with energy density (with $\overline{\varphi}=\varphi^+\gamma^0$)
\begin{eqnarray}
\mathcal{H}_{\varphi}&=&mc^2\overline{\varphi}\cdot\varphi
-\frac{\mathrm{i}}{2}\hbar c\overline{\varphi}\cdot\bm{\gamma}\cdot
\bm{\nabla}\varphi
+\frac{\mathrm{i}}{2}\hbar c\bm{\nabla}\overline{\varphi}\cdot\bm{\gamma}\cdot
\varphi.
  \end{eqnarray}
The Dirac field (\ref{eq:diracfs1},\ref{eq:1pdirac}) would also carry momentum
\begin{eqnarray}\nonumber
\bm{p}&=&\int\!d^3\bm{x}\,\frac{\hbar}{2\mathrm{i}}\left[\varphi^+(\bm{x},t)\cdot
\bm{\nabla}\varphi(\bm{x},t)
-\bm{\nabla}\varphi^+(\bm{x},t)\cdot\varphi(\bm{x},t)\right]
\\
&=&\int\!d^3\bm{k}\,\hbar\bm{k}\sum_s|\psi_s(\bm{k})|^2,
  \end{eqnarray}
and charge
\begin{eqnarray} \label{eq:charge}
  Q&=&q\int\!d^3\bm{x}\,\varphi^+(\bm{x},t)\cdot\varphi(\bm{x},t)
  =q\int\!d^3\bm{k}\sum_s|\psi_s(\bm{k})|^2=q.
  \end{eqnarray}
Eq.~(\ref{eq:charge}) implies in particular that single-particle normalization
of the modes $\psi_s(\bm{k})$ or $\chi_s(\bm{k})$, respectively, implies
normalized Dirac spinors $\varphi(\bm{x},t)$, and this leads to the identification
of the normalized charge
density $\varrho(\bm{x},t)/q=\varphi^+(\bm{x},t)\cdot\varphi(\bm{x},t)$
with a Born probability density for
particle position, which also leads to the equation for the
Heisenberg picture velocity
operator \cite{greiner},
\begin{equation}
\dot{\bm{x}}=\frac{\mathrm{i}}{\hbar}[\gamma^0(mc^2+c\bm{\gamma}\cdot
\bm{p}),\bm{x}]=c\gamma^0\bm{\gamma}.
\end{equation}
On the other hand, the Klein--Gordon field (with $k\cdot x\equiv\bm{k}\cdot\bm{x}-\omega_{\bm{k}}t$)
\begin{eqnarray} \label{eq:phiKG}
\phi(\bm{x},t)&=&\frac{1}{\sqrt{2\pi}^3}
\int\frac{d^3\bm{k}}{\sqrt{2\omega_{\bm{k}}}}\Big[\psi(\bm{k})
\exp(\mathrm{i}k\cdot x)+\chi^+(\bm{k})
\exp(-\,\mathrm{i}k\cdot x)\Big],
\end{eqnarray}
with single-particle normalization,
\begin{equation}\label{eq:1pKG}
\int\!d^3\bm{k}\,|\psi(\bm{k})|^2=1,\quad\chi(\bm{k})=0,
\end{equation}
satisfies
\begin{eqnarray}\nonumber
  Q&=&\mathrm{i}q\int\!d^3\bm{x}\left[\phi^+(\bm{x},t)\dot{\phi}(\bm{x},t)
    -\dot{\phi}^+(\bm{x},t)\phi(\bm{x},t)\right]
  \\ \label{eq:chargeKG}
&=&q\int\!d^3\bm{k}\,|\psi(\bm{k})|^2=q.
\end{eqnarray}
However, in this case the normalized charge density
\begin{equation}
\varrho_\phi(\bm{x},t)/q=\mathrm{i}\left[\phi^+(\bm{x},t)\dot{\phi}(\bm{x},t)
    -\dot{\phi}^+(\bm{x},t)\phi(\bm{x},t)\right]
  \end{equation}
is not positive definite, not even in the single-particle case (\ref{eq:1pKG}),
and therefore $\varrho_\phi(\bm{x},t)/q$ is not suitable to define a probability density for
particle location. We will see an explicit example of negative regions of the
normalized charge density for the two-dimensional single-particle Klein--Gordon field
in Fig.~\ref{fig:rhoKG0}. 

On the other hand, the energy density of the Klein--Gordon field,
\begin{eqnarray}\nonumber
  \mathcal{H}_{\phi}(\bm{x},t)&=&\frac{m^2c^4}{\hbar}\left|\phi(\bm{x},t)\right|^2
  +\hbar\dot{\phi}^+(\bm{x},t)\cdot\dot{\phi}(\bm{x},t)
\\ \label{eq:KGEdensity}
&&+\hbar c^2\bm{\nabla}\phi^+(\bm{x},t)\cdot\bm{\nabla}\phi(\bm{x},t),
\end{eqnarray}
\begin{eqnarray}\nonumber
  E&=&\int\!d^3\bm{x}\,\mathcal{H}_{\phi}(\bm{x},t),
\end{eqnarray}
is positive definite. Therefore it is tempting to use the normalized energy
density $\mathcal{H}_{\phi}(\bm{x},t)/E$ as a probability measure for scalar particle
location if the single-particle conditions (\ref{eq:1pKG}) are fulfilled \cite{bialynicki2}. 

A similar reasoning applies to photons, which have no charge density to start with
but a positive energy density
\begin{equation}
  \mathcal{H}_{\gamma}(\bm{x},t)=
  \frac{\epsilon_0}{2}\bm{E}^2(\bm{x},t)+\frac{1}{2\mu_0}\bm{B}^2(\bm{x},t).
\end{equation}
In this case, consideration of $\mathcal{H}_{\gamma}/E$ as a probability measure for
photon location leads to electromagnetic fields as natural candidates for
photon wave functions 
\cite{oppenheimer,bialynicki1,sipe,bialynicki1b,scully,keller1,keller2,raymer,BBBB1,BBBB2},
e.g.~in the form of a normalized Riemann--Silberstein vector
$(\sqrt{\epsilon_0/2}\bm{E}+\mathrm{i}\bm{B}/\sqrt{2\mu_0})/\sqrt{E}$.

An attractive feature of the normalized energy
density $\mathcal{H}(\bm{x},t)/E$ as a
probability measure for signal location concerns the resulting
velocity equation. The equation
\begin{equation}\label{eq:expectx}
\langle\bm{x}\rangle(t)=\frac{1}{E}\int\!d^3\bm{x}\,\bm{x}\mathcal{H}(\bm{x},t)
\end{equation}
yields for the velocity of any field configuration
with energy density $\mathcal{H}(\bm{x},t)=T^{00}(\bm{x},t)$
and energy current density $\bm{\mathcal{S}}(\bm{x},t)$
the correct velocity relation $\bm{v}=c^2\bm{p}/E$. This is a consequence
of energy conservation,
\begin{eqnarray}
  \partial_t\mathcal{H}(\bm{x},t)&=&c\partial_0 T^{00}(\bm{x},t)
  =-\,c\partial_iT^{0i}(\bm{x},t)
  =-\,\bm{\nabla}\cdot\bm{\mathcal{S}}(\bm{x},t),
\end{eqnarray}
and symmetry of the energy-momentum
tensor, $T^{0i}(\bm{x},t)=T^{i0}(\bm{x},t)=c\mathcal{P}^i(\bm{x},t)$,
which relates energy current density and momentum density
\begin{equation}
  \bm{\mathcal{S}}(\bm{x},t)=c^2\bm{\mathcal{P}}(\bm{x},t).
\end{equation}
Eq.~(\ref{eq:expectx}) therefore yields
\begin{equation}\label{eq:expectv}
  \frac{d}{dt}\langle\bm{x}\rangle(t)
  =\frac{1}{E}\int\!d^3\bm{x}\,\bm{\mathcal{S}}(\bm{x},t)
  =\frac{c^2\bm{p}}{E}.
\end{equation}

Unfortunately, contrary to the bosonic cases, the normalized energy
density $\mathcal{H}_{\varphi}(\bm{x},t)/E$ of the Dirac field cannot serve
as a probability measure for fermion location,
because even in the single-particle case (\ref{eq:1pdirac}) this measure can become
negative. We will see a two-dimensional example for negative regions of
the normalized single-fermion energy density in Fig.~\ref{fig:NormTildeH0},
as explained in Sec.~\ref{sec:dirac}.

We therefore find a schism between fermions and bosons in terms
of suggested probability measures for relativistic particle location:
 Fermions seem to have a probability measure on the basis of the normalized
charge density $\varrho(\bm{x},t)/q$ for single-particle amplitudes (\ref{eq:1pdirac}),
whereas the dominant proposals for bosons are based on normalized
energy densities $\mathcal{H}(\bm{x},t)/E$ for single-particle amplitudes
(\ref{eq:1pKG}). 

 For yet another possibility, the $\bm{k}$-space modes $\psi_s(\bm{k})$ for free
scalar, spin-$1/2$, and spin-1 fields all yield energy and momentum expectation
values of the same kind if we calculate expectation values for
single-particle states
\begin{equation}\label{eq:psistatek}
  \bm{|}\psi(t)\bm{\rangle}=\int\!d^3\bm{k}\sum_s a_s^+(\bm{k})\bm{|}0\bm{\rangle}
  \psi_s(\bm{k})\exp(-\,\mathrm{i}\omega_{\bm{k}}t),
\end{equation}
\begin{equation}
  \int\!d^3\bm{k}\sum_s |\psi_s(\bm{k})|^2=1,
\end{equation}
\textit{viz.}
\begin{eqnarray}\nonumber
  &&H=\int\!d^3\bm{k}\sum_s\hbar\omega_{\bm{k}}\,a_s^+(\bm{k})a_s(\bm{k})
  \\ \label{eq:Epsi}
  &&\to
  \bm{\langle}\psi(t)\bm{|}H\bm{|}\psi(t)\bm{\rangle}
  =\int\!d^3\bm{k}\sum_s\hbar\omega_{\bm{k}}\,\psi_s^+(\bm{k})\psi_s(\bm{k}),
\end{eqnarray}
\begin{eqnarray}\nonumber
  &&\bm{P}=\int\!d^3\bm{k}\sum_s\hbar\bm{k}\,a_s^+(\bm{k})a_s(\bm{k})
  \\
  &&\to \bm{\langle}\psi(t)\bm{|}\bm{P}\bm{|}\psi(t)\bm{\rangle}
  =\int\!d^3\bm{k}\sum_s\hbar\bm{k}\,\psi_s^+(\bm{k})\psi_s(\bm{k}),
\end{eqnarray}
\begin{eqnarray}\nonumber
  &&Q=q\int\!d^3\bm{k}\sum_sa_s^+(\bm{k})a_s(\bm{k})
  \\ \label{eq:Qpsi}
  &&\to
  \bm{\langle}\psi(t)\bm{|}Q\bm{|}\psi(t)\bm{\rangle}
  =q\int\!d^3\bm{k}\sum_s\psi_s^+(\bm{k})\psi_s(\bm{k}).
\end{eqnarray}
They also yield the correct expectation values for the spin-operators
(where applicable),
\begin{eqnarray} \label{eq:Spsi}
  &&S_3=\int\!d^3\bm{k}\sum_{ss'}\hbar\,a_s^+(\bm{k})(\Sigma_3)_{ss'}a_{s'}
  (\bm{k})
  \to
  \bm{\langle}\psi(t)\bm{|}S_3\bm{|}\psi(t)\bm{\rangle}
  ={\hbar}s,
\end{eqnarray}
where $\Sigma_3$ is the
Pauli matrix $\sigma_3/2$ for spin $1/2$, $s\in\{1/2,-\,1/2\}$, or
$\Sigma_3=\mathrm{diag}(1,0,-\,1)$ for spin 1, $s\in\{1,0,-\,1\}$.

These equations clearly imply that the function $\psi_s(\bm{k})$ is
a $\bm{k}$-space probability amplitude to find a particle with
momentum $\hbar\bm{k}$, energy $\hbar\omega_{\bm{k}}$, charge $q$ (where applicable),
and spin projection
$\hbar s$ (where applicable), and they hold irrespective
of the magnitude of $|\bm{k}|$, i.e.~we do have well-defined normalizable
$\bm{k}$-space wave functions also in the ultrarelativistic limit. However,
Parseval's theorem implies that the Fourier transform
\begin{equation}\label{eq:psisxt}
  \psi_s(\bm{x},t)=\frac{1}{\sqrt{2\pi}^3}\int\!d^3\bm{k}\,\psi_s(\bm{k})
    \exp[\mathrm{i}(\bm{k}\cdot\bm{x}-\omega_{\bm{k}}t)]
\end{equation}
also provides normalized scalars, spinors or vectors, respectively,
and the experience with nonrelativistic quantum mechanics
would make us expect that $\psi_s(\bm{x},t)$ should serve as a particle wave packet
in $\bm{x}$ space if $\psi_s(\bm{k},t)=\psi_s(\bm{k})\exp(-\,\mathrm{i}\omega_{\bm{k}}t)$
is a particle wave packet
in $\bm{k}$ space. Further support for this comes from the Fourier transformation
of particle creation operators,
\begin{equation}
  a_s^+(\bm{x})=\frac{1}{\sqrt{2\pi}^3}\int\!d^3\bm{k}\,a_s^+(\bm{k})
    \exp(-\,\mathrm{i}\bm{k}\cdot\bm{x}).
  \end{equation}
Eqs.~(\ref{eq:psistatek}) and (\ref{eq:psisxt}) then imply
\begin{equation}\label{eq:psistatex}
  \bm{|}\psi(t)\bm{\rangle}=\int\!d^3\bm{x}\sum_s a_s^+(\bm{x})\bm{|}0\bm{\rangle}
  \psi_s(\bm{x},t),
\end{equation}
whence $\psi_s(\bm{x},t)$ would appear to be a single-particle creation amplitude
in direct space in the same vein as $\psi_s(\bm{k},t)$ is
a single-particle creation amplitude in wave-vector space.

We test this intuitive expectation through numerical
evaluations of Gaussian wave packets and their related quantum fields,
as well as evaluation of their canonical energy densities and the related energy
pseudo-densities
\begin{eqnarray}
  \tilde{\mathcal{H}}(\bm{x},t)&=&\frac{\mathrm{i}\hbar}{2}\sum_s
  \left(\psi_s^+(\bm{x},t)\cdot\partial_t\psi_s(\bm{x},t)
  -\partial_t\psi_s^+(\bm{x},t)\cdot\psi_s(\bm{x},t)\right),
\end{eqnarray}
which yield the same energy expectation values (\ref{eq:Epsi})
as the canonical energy densities
of relativistic fields upon spatial integration.

Stated differently, we study the following question:
If the amplitude $\psi_s(\bm{x},t)$ is not an acceptable
wave function in terms of the Born interpretation, 
\textit{how far away from the normalized canonical energy
  density $\mathcal{H}(\bm{x},t)/E$
  are $\sum_s|\psi_s(\bm{x},t)|^2$ and $\tilde{\mathcal{H}}(\bm{x},t)/E$ actually?}

As a preparation for the discussion of the possibility of nonlocal relations
between relativistic wave functions and quantum fields,
we will start our investigation with a discussion of an aspect of nonrelativistic
quantum mechanics:
Sec.~\ref{sec:nonrel} emphasizes the
distinction between the Born probability
density $|\psi|^2$ for particle location on the one hand,
and the canonical energy
density $\mathcal{H}=(\hbar^2/2m)\bm{\nabla}\psi^+\cdot\bm{\nabla}\psi+\psi^+V\psi$
on the other hand. The manifest difference of the two densities can lead to
macroscopic separation of signals between a particle detector in the Born sense
versus a detector that would track the disposable energy of a nonrelativistic particle.

The observations from Sec.~\ref{sec:nonrel} motivate us
in Secs.~\ref{sec:scalar}--\ref{sec:scalar2} 
to assume a devil's advocate position and argue for a nonlocal relation
between relativistic wave functions $\psi(\bm{x},t)$ and the
corresponding first-quantized Klein--Gordon fields $\phi(\bm{x},t)$ (which are related
to quantum Klein--Gordon fields $\Phi(\bm{x},t)$ e.g.~through expectation values for
coherent states).
It is important to explore this possibility: The use of spatial Fourier
transforms $\psi(\bm{x},t)$ of the normal modes of
a quantum field $\phi(\bm{x},t)$ has many
attractive features, including standard normalizability, a standard
momentum-position uncertainty relation if $|\psi(\bm{x},t)|^2$ could be
adopted as a Born measure for particle location, and local expressions for
energy-momentum densities.

Sec.~\ref{sec:latency} explores the origin 
of a latency effect that we find in $|\psi(\bm{x},t)|^2$
and in two other proxies for particle
position, and Sec.~\ref{sec:mass} discusses the impact of mass and momentum
of the Klein--Gordon field.

Sec.~\ref{sec:dirac} translates our results into implications for the Dirac field.

We summarize in Sec.~\ref{sec:conc} and confirm that, within a local description
of relativistic dynamics, normalized charge density
provides the best
possible proxy for fermion position, whereas normalized canonical energy
density provides the best proxy for boson position.

  \section{Separation of disposable particle energy and particle position
    in nonrelativistic quantum mechanics\label{sec:nonrel}}

  Nonrelativistic quantum mechanics appears both local and in agreement with the Born
  interpretation because the probability density to find a particle in a location $\bm{x}$
  at time $t$,
  \begin{equation}\label{eq:rhonr}
    \varrho(\bm{x},t)=|\psi(\bm{x},t)|^2,
  \end{equation}
  the disposable energy density
  \begin{eqnarray}\label{eq:Hnonrel}
    \mathcal{H}(\bm{x},t)&=&
    \frac{\hbar^2}{2m}\bm{\nabla}\psi^+(\bm{x},t)\cdot\bm{\nabla}\psi(\bm{x},t)
    +\psi^+(\bm{x},t)V(\bm{x})\psi(\bm{x},t),
  \end{eqnarray}
  the momentum density
  \begin{eqnarray} \label{eq:mathcalP}
    \bm{\mathcal{P}}(\bm{x},t)&=&\frac{\hbar}{2\mathrm{i}}\left[
      \psi^+(\bm{x},t)\cdot\bm{\nabla}\psi(\bm{x},t)
      -\bm{\nabla}\psi^+(\bm{x},t)\cdot\psi(\bm{x},t)\right],
  \end{eqnarray}
  the angular momentum density
  \begin{equation} \label{eq:mathcalM}
    \bm{\mathcal{M}}(\bm{x},t)=\bm{x}\times\bm{\mathcal{P}}(\bm{x},t),
    \end{equation}
  and also the spin density (where applicable,
  e.g.~$\underline{\bm{S}}=\hbar\underline{\bm{\sigma}}/2$ for spin-1/2 particles)
  \begin{equation}\label{eq:mathcalS}
    \bm{S}(\bm{x},t)
    =\psi^+(\bm{x},t)\cdot\underline{\bm{S}}\cdot\psi(\bm{x},t)
    \end{equation}
  are local expressions in terms of the wave function $\psi(\bm{x},t)$. Furthermore,
  the probability density $\varrho(\bm{x},t)$ satisfies the local conservation law
  \begin{equation}
\frac{\partial}{\partial t}\varrho(\bm{x},t)=-\,\bm{\nabla}\cdot\bm{\mathcal{P}}(\bm{x},t)/m,
  \end{equation}
  such that the velocity density $\bm{v}(\bm{x},t)=\bm{\mathcal{P}}(\bm{x},t)/m$
  serves as a probability current density.
 
  On the face of it, the mere fact that all the
  densities (\ref{eq:rhonr}-\ref{eq:mathcalS}) of basic
  particle observables are given as local expressions in terms of the
  wave function $\psi(\bm{x},t)$ make nonrelativistic quantum mechanics certainly \textit{look}
  like a local theory, and if we define ``local theory'' as a theory where all densities for
  observables are given in terms of local expression of a wave function, or at least
  in terms of local expressions with respect to a quantum field, then nonrelativistic quantum
  mechanics is a local theory by definition. However, one might ask:
  Does locality in terms of local expressions with respect to wave functions
  or quantum fields also imply physical locality in the sense that
  densities of observables (e.g.~the energy density (\ref{eq:Hnonrel})) are closely
  correlated with particle probability densities? Basic examples demonstrate that the answer
  to this question is ``No''.

\begin{figure}[htb]\begin{center}
\scalebox{1}{\includegraphics{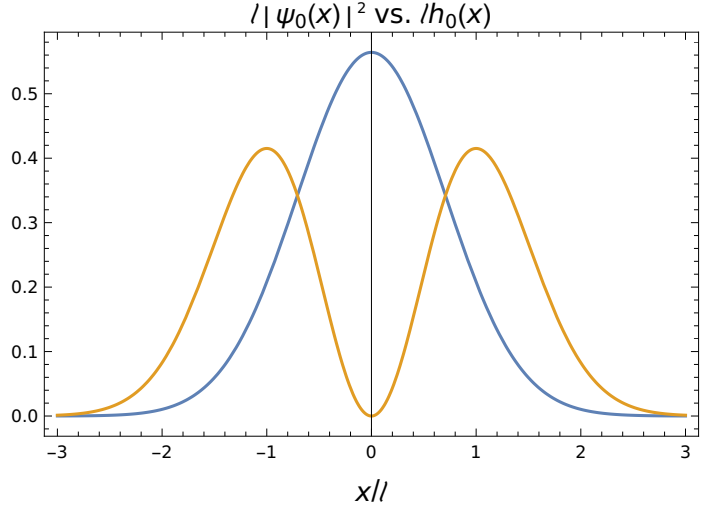}}
\caption{\label{fig:CompRhoH0}
  The Born probability density $|\psi_0({x})|^2$ (blue) and the normalized energy
  density $h_0({x})$ (orange) for the ground state of the harmonic oscillator.
The length unit is $\ell=\sqrt{\hbar/m\omega}$.}
\end{center}\end{figure}

\begin{figure}[htb]\begin{center}
\scalebox{1}{\includegraphics{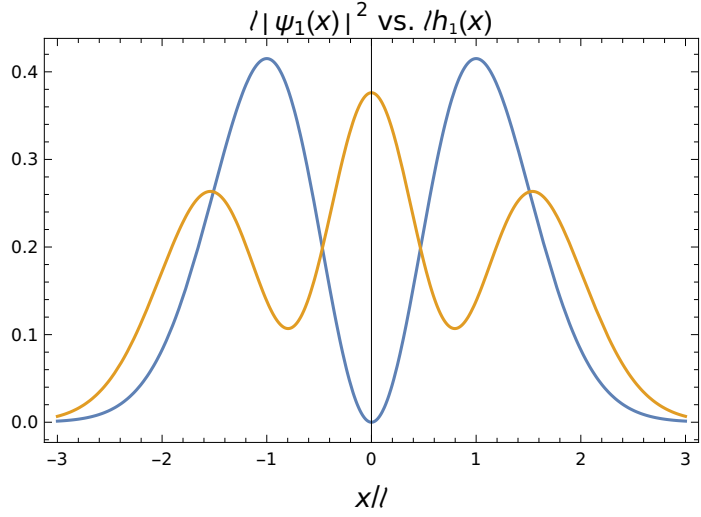}}
\caption{\label{fig:CompRhoH1}
  The Born probability density $|\psi_1({x})|^2$ (blue) and the normalized energy
  density $h_1({x})$ (orange) for the first excited state of the oscillator.
  The length unit is $\ell=\sqrt{\hbar/m\omega}$.}
 \end{center}\end{figure}

  For example, we plot the probability density (\ref{eq:rhonr}) and the normalized
  energy density $h(x,t)=\mathcal{H}(x,t)/\int_{-\infty}^\infty\!dy\,\mathcal{H}(y,t)$ for
  oscillator eigenstates,
  \begin{equation}\label{eq:Hnorm}
  h_n({x})=\left.\frac{2}{2n+1}\frac{\mathcal{H}({x})}{\hbar\omega}\right|_{\psi=\psi_n},
     \end{equation}
  in Figs.~\ref{fig:CompRhoH0} and \ref{fig:CompRhoH1}.
  If a particle detection experiment in a cold sample of oscillators would not detect
  the probability density $|\psi_0({x})|^2$, but the nonrelativistic energy
  density (\ref{eq:Hnonrel}) of the particles, sampling many observations should produce
  the orange curve in Fig.~\ref{fig:CompRhoH0}, but not the blue curve which is
  predicted by the Born interpretation. For electrons oscillating with a frequency $f=727$ Hz
  this yields a macroscopic separation of $\ell=1$ mm between the single bright fringe predicted
  by the Born interpretation and the two bright fringes that we might
  intuitively expect from the electrons' energy densities.
  
 There are two different possibilities to make the normalized energy
 density $h(\bm{x})$ for an energy eigenstate with wave function $\psi(\bm{x})$
 coincide with the Born probability density $|\psi(\bm{x})|^2$. On the one hand, we can
 replace the hermitian Hamiltonian density (\ref{eq:Hnonrel}) with another hermitian density
  \begin{eqnarray}\nonumber
    \tilde{\mathcal{H}}(\bm{x},t)&=&\mathcal{H}(\bm{x},t)-\frac{\hbar^2}{4m}
    \bm{\nabla}\cdot[\psi^+(\bm{x},t)\bm{\nabla}\psi(\bm{x},t)]
    \\ \nonumber
    &&-\frac{\hbar^2}{4m}
    \bm{\nabla}\cdot[\bm{\nabla}\psi^+(\bm{x},t)\cdot\psi(\bm{x},t)]
    \\ \nonumber
    &=&\psi^+(\bm{x},t)V(\bm{x})\psi(\bm{x},t)-\frac{\hbar^2}{4m}\psi^+(\bm{x},t)
       \Delta\psi(\bm{x},t)
    \\ \label{eq:Hnonrel2}
        &&-\frac{\hbar^2}{4m}\Delta\psi^+(\bm{x},t)\cdot\psi(\bm{x},t).
  \end{eqnarray}
  For energy eigenstates, this yields $h_n(\bm{x})\to\tilde{h}_n(\bm{x})=|\psi_n(\bm{x})|^2$.

  The transformation (\ref{eq:Hnonrel2}) also changes the local energy conservation law from
  \begin{equation}
  \frac{\partial}{\partial t}\mathcal{H}(\bm{x},t)=-\,\bm{\nabla}\cdot\bm{J}(\bm{x},t),
  \end{equation}
  with the energy current density
  \begin{eqnarray}\label{eq:jHnonrel}
    \bm{J}&=&-\,\frac{\hbar^2}{2m}\left(
      \frac{\partial\psi^+}{\partial t}\cdot\bm{\nabla}\psi
      +\bm{\nabla}\psi^+\cdot\frac{\partial\psi}{\partial t}\right)\!,
  \end{eqnarray}
to
 \begin{equation}
   \frac{\partial}{\partial t}\tilde{\mathcal{H}}(\bm{x},t)=-\,\bm{\nabla}\cdot
   \tilde{\bm{J}}(\bm{x},t),
  \end{equation}
  with the energy current density
  \begin{eqnarray}\nonumber
    \tilde{\bm{J}}&=&\bm{J}+\frac{\hbar^2}{4m}\frac{\partial}{\partial t}
    \left(\psi^+\cdot\bm{\nabla}\psi
      +\bm{\nabla}\psi^+\cdot\psi\right)
    \\ \label{eq:jHnonrel2} 
    &=&\frac{\hbar^2}{4m}\left(\psi^+\cdot
      \frac{\partial}{\partial t}\bm{\nabla}\psi
      -\frac{\partial\psi^+}{\partial t}\cdot\bm{\nabla}\psi
     +\frac{\partial}{\partial t}\bm{\nabla}\psi^+\cdot\psi
      -\bm{\nabla}\psi^+\cdot\frac{\partial\psi}{\partial t}\right)\!.
  \end{eqnarray}

  We also note that on-shell, the alternative Hamiltonian density $\tilde{\mathcal{H}}(\bm{x},t)$
  corresponds to a four-dimensional extension of the momentum density $\bm{\mathcal{P}}$
  (\ref{eq:mathcalP}) in the sense that
  \begin{eqnarray}\nonumber
    \mathcal{P}_0(\bm{x},t)&=&-\,\tilde{\mathcal{H}}(\bm{x},t)/c
    \\
    &=&\frac{\hbar}{2\mathrm{i}}\left[\psi^+(\bm{x},t)\cdot\partial_0\psi(\bm{x},t)
    -\partial_0\psi^+(\bm{x},t)\cdot\psi(\bm{x},t)\right]\!.
    \end{eqnarray}

  The transformation (\ref{eq:Hnonrel2}) of Hamiltonian densities implies for the
  Lagrange density of the Schr\"odinger field that we replace
  \begin{eqnarray} \label{eq:Lpsi}
\mathcal{L}&=&\frac{\mathrm{i}\hbar}{2}
\left(\psi^+\cdot\frac{\partial\psi}{\partial t}
-\frac{\partial\psi^+}{\partial t}\cdot\psi\right)
-\frac{\hbar^2}{2m}\bm{\nabla}\psi^+\cdot\bm{\nabla}\psi
-\psi^+ V\psi
\end{eqnarray}
with
  \begin{eqnarray} \label{eq:Lpsi2}
\tilde{\mathcal{L}}&=&\frac{\mathrm{i}\hbar}{2}
\left(\psi^+\cdot\frac{\partial\psi}{\partial t}
-\frac{\partial\psi^+}{\partial t}\cdot\psi\right)
+\frac{\hbar^2}{4m}\left(\psi^+\Delta\psi
+\Delta\psi^+\cdot\psi\right)
-\psi^+ V\psi.
\end{eqnarray}

\begin{figure}[htb]\begin{center}
\scalebox{1}{\includegraphics{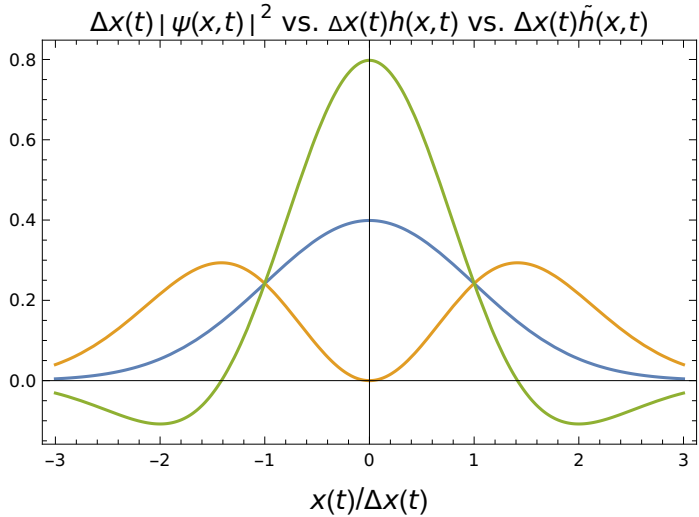}}
\caption{\label{fig:FreeGauss}
  The Born probability density $|\psi({x},t)|^2$ (blue), the normalized energy
  density $h({x},t)$ (orange) and the modified normalized energy
  density $\tilde{h}({x},t)$ (green) for the freely evolving Gaussian package (\ref{eq:gausst}).}
\end{center}\end{figure}

  The change $\mathcal{L}\to\tilde{\mathcal{L}}$ changes $\mathcal{H}\to\tilde{\mathcal{H}}$
  but preserves the momentum density $\bm{\mathcal{P}}$ (\ref{eq:mathcalP})
  and also the densities (\ref{eq:mathcalM},\ref{eq:mathcalS}). Rationalizing the Born
  probability density as a normalized energy density seems attractive, but there are
  caveats:\\
  -- Both of the normalized energy densities $h(\bm{x})$ and $\tilde{h}(\bm{x})$ can have
  negative components if there are regions where $V(\bm{x})<0$. The density $\tilde{h}(\bm{x})$
  becomes negative even for free particle wave packets, see Fig.~\ref{fig:FreeGauss},
  where the different densities are displayed for a nonrelativistic Gaussian package at rest,
\begin{eqnarray} \nonumber
\psi(x,t)&=&\frac{(2\pi\Delta x^2)^{1/4}}{\left[2\pi\Delta x^2
+\mathrm{i}\pi(\hbar t/m)\right]^{1/2}}
\exp\!\left(-\,\frac{x^2}{4\Delta x^2+
(\hbar^2 t^2/m^2\Delta x^2)}
\right)
\\ \label{eq:gausst}
&&\times
\exp\!\left(
\mathrm{i}\frac{\hbar t}{8m}
\frac{x^2}{(\Delta x^2)^2+
(\hbar^2 t^2/4m^2)}
\right)\!,
\end{eqnarray}
which has constant width $\Delta k=1/2\Delta x$ in $k$ space
but spreads in $x$ space according to
\begin{equation}\label{eq:dx2t}
\Delta x^2(t)=\Delta x^2+\frac{\hbar^2 t^2}{4m^2\Delta x^2},
\end{equation}
see e.g.~\cite{rf:rdick}.

  \noindent
  -- Furthermore, the true energy-momentum tensor of a theory is defined through variation
  with respect to the spacetime metric, and this yields the first-order expression
  (\ref{eq:Hnonrel}). Variation with respect to the metric is not affected through
  addition of complete derivative terms to the Lagrange density. The effect of the
  second order derivatives in $\tilde{\mathcal{L}}$ would only amount to addition of
  Christoffel symbols which reduce the $T^{00}$ component of the energy-momentum tensor
  to the first order expression $\mathcal{H}$ (\ref{eq:Hnonrel}) upon variation.
  Agreement of the gravitational energy-momentum tensor with the canonical energy-momentum
  tensor from time-translation invariance therefore leads us back to the standard
  densities $\mathcal{L}$ and $\mathcal{H}$.

  Another option to make the normalized energy density and the Born probability
  density coincide in nonrelativistic quantum mechanics
  results from the observation that the actual nonrelativistic limit
  of relativistic Hamiltonians also includes the rest energy,
  \begin{equation}
    \hat{\mathcal{H}}(\bm{x},t)=\mathcal{H}(\bm{x},t)+mc^2|\psi(\bm{x},t)|^2.
  \end{equation}
  This yields $\hat{h}(\bm{x},t)\to |\psi(\bm{x},t)|^2$ in the nonrelativistic limit.
 
  There are two lessons from these basic considerations:\\
  -- The maxima of the Born particle probability density on the one hand and the
  disposable energy density on the other hand can be macroscopically separated in
  nonrelativistic quantum mechanics: The mathematical locality of Eq.~(\ref{eq:Hnonrel})
  does not imply physical locality.\\
  -- The normalized density of the disposable energy of a particle cannot replace the
  Born density $|\psi(\bm{x},t)|^2$ as a probability density for particle location.
  
  On the face of it, these observations would seem to require that wave functions
  also exist beyond the nonrelativistic limit of quantum theory.
We will see in the following section that the Fourier transforms $\psi(\bm{x},t)$
of the normal mode
amplitudes\footnote{
  The normal mode is $\exp[\mathrm{i}(\bm{k}\cdot\bm{x}-\omega_{\bm{k}}t)]/\sqrt{2\pi}^3$,
  and with the extraction of the factor $1/\sqrt{2\omega_{\bm{k}}}$ for simple commutation
  relations,
  the normal mode amplitude in the quantum field $\Phi(x)$ is $a(\bm{k})$,
  or the corresponding function $\psi(\bm{k})$ in the first-quantized field $\phi(x)$.
  However, for brevity, we also
  denote $\psi(\bm{k},t)=\psi(\bm{k})\exp(-\,\mathrm{i}\omega_{\bm{k}}t)$
  as normal mode amplitudes.}
 $\psi(\bm{k},t)=\psi(\bm{k})\exp(-\,\mathrm{i}\omega_{\bm{k}}t)$ 
of first-quantized relativistic fields $\phi(\bm{x},t)$ have properties that we would 
expect from wave functions. However, the functions $\psi(\bm{x},t)$
 yield local $\bm{x}$-space expressions
 for energy, momentum and charge densities which differ from the corresponding
 canonical
expressions in terms of the quantum fields. Indeed, the different expressions for
local densities always yield the same $\bm{k}$-space densities, but they 
coincide in $\bm{x}$ space only in the nonrelativistic limit or for small
momentum uncertainy $\Delta p\ll p$.
We will explore the relations between the different $\bm{x}$-space densities
in Sec.~\ref{sec:scalar2} to understand when the functions $\psi(\bm{x},t)$
could provide relativistic wave functions.

\section{Normal modes as relativistic wave functions for scalar particles?\label{sec:scalar}}

We consider a complex scalar quantum field $\Phi(x)$ that can locally couple to 
other fields
through Yukawa and gauge couplings. To be specific, we include electromagnetic 
interactions and a Yukawa self-interaction in the Lagrange density,
\begin{eqnarray}\nonumber
\mathcal{L}&=&-\,\hbar c^2\left(\partial_\mu\Phi^++\mathrm{i}\frac{q}{\hbar}\Phi^+A_\mu\right)
\!\left(\partial^\mu\Phi-\mathrm{i}\frac{q}{\hbar}A^\mu\Phi\right)
\\ \label{eq:Lscalar}
&&-\frac{m^2 c^4}{\hbar}\Phi^+\Phi-\hbar c^3\frac{\lambda}{4}(\Phi^+\Phi)^2
-\frac{1}{4\mu_0}F_{\mu\nu}F^{\mu\nu}.
\end{eqnarray}
The time $(t)$ and length $(\ell)$ dimensions of the scalar field are $t^{1/2}\ell^{-3/2}$
such that no extra constants
appear in the free mode expansion (\ref{eq:KGsolve}) and the modes $a(\bm{k})$
and $b(\bm{k})$ have dimensions $\ell^{3/2}$.

In spite of the interactions, the quanta of the
scalar field are determined through the Fourier decomposition of the  
quantum field in the interaction picture,
\begin{eqnarray}\nonumber
\Phi(x)&=&\frac{1}{\sqrt{2\pi}^3}
\int\frac{d^3\bm{k}}{\sqrt{2\omega_{\bm{k}}}}\Big(
a(\bm{k})\exp\!\left[\mathrm{i}(\bm{k}\cdot\bm{x}
-\omega_{\bm{k}}t)\right]
\\ \label{eq:KGsolve}
&&+b^+(\bm{k})\exp\!\left[-\,\mathrm{i}(\bm{k}\cdot\bm{x}
-\omega_{\bm{k}}t)\right]\Big),
\end{eqnarray}
where $\omega_{\bm{k}}$ is given by
\begin{equation}\label{eq:defomegafree}
\omega_{\bm{k}}=c\sqrt{\bm{k}^2+(mc/\hbar)^2}.
\end{equation}
The interaction picture Hamiltonian,
\begin{eqnarray}\nonumber
  H_I&=&\int\!d^3\bm{x}\,\bigg(\mathrm{i}qc^2\left(\Phi^+\bm{A}\cdot\bm{\nabla}\Phi
    -\bm{\nabla}\Phi^+\cdot\bm{A}\Phi\right)
    \\ \nonumber
    &&+\mathrm{i}qc\left(\dot{\Phi}^+A^0\Phi
    -\Phi^+A^0\dot{\Phi}\right)+\hbar c^3\frac{\lambda}{4}(\Phi^+\Phi)^2
    \\ \label{eq:HI}
    &&+\frac{q^2}{\hbar}c^2\Phi^+\left[(A^0)^2+\bm{A}^2\right]\Phi
    \bigg),
\end{eqnarray}
contains only the freely evolving quantum fields
of the interaction picture and describes
scattering of the scalar (anti-)particles through the scattering matrix.

A general single-particle state of the scalar field has the form
\begin{eqnarray} \nonumber
\bm{|}\psi(t)\bm{\rangle}
&=&\int\!d^3\bm{k}\,a^+(\bm{k})\bm{|}0\bm{\rangle}\psi(\bm{k})
\exp(-\,\mathrm{i}\omega_{\bm{k}}t)
\\ \label{eq:1p}
&=&\int\!d^3\bm{x}\,a^+(\bm{x})\bm{|}0\bm{\rangle}\psi(\bm{x},t),
\end{eqnarray}
with the normal mode operators in $\bm{x}$ space,
\begin{equation}
a(\bm{x})=\frac{1}{\sqrt{2\pi}^3}\int\!d^3\bm{k}\,a(\bm{k})
\exp(\mathrm{i}\bm{k}\cdot\bm{x}).
\end{equation}
The wave packet in $\bm{x}$ space is related to the wave packet in 
$\bm{k}$ space through
\begin{equation}\label{eq:varphixt}
\psi(\bm{x},t)=\frac{1}{\sqrt{2\pi}^3}\int\!d^3\bm{k}\,\psi(\bm{k})
\exp\!\left[\mathrm{i}(\bm{k}\cdot\bm{x}
-\omega_{\bm{k}}t)\right].
\end{equation}

Single-particle normalization of the $\bm{k}$-space wave packet
implies single-particle normalization of the corresponding $\bm{x}$-space 
wave packet,
\begin{eqnarray}\label{eq:normphi1}
  \int\!d^3\bm{k}\left|\psi(\bm{k})\right|^2&=&
  \int\!d^3\bm{x}\left|\psi(\bm{x},t)\right|^2=1.
\end{eqnarray}

The $\bm{k}$ space wave packet is also a normal mode of the first quantized
scalar field (here $\phi(x)\equiv\phi(\bm{x},t)$, i.e.~$x$ denotes
the 4-vector $(ct,\bm{x})$, and $\Phi(\bm{x})$ is the Klein--Gordon
quantum field in the
Schr\"odinger picture),
\begin{eqnarray}\nonumber
\phi(x)&=&\bm{\langle}0\bm{|}\Phi(x)\bm{|}\psi(0)\bm{\rangle}
=\bm{\langle}0\bm{|}\Phi(\bm{x})\bm{|}\psi(t)\bm{\rangle}
\\ \label{eq:phic1}
&=&\frac{1}{\sqrt{2\pi}^3}
\int\frac{d^3\bm{k}}{\sqrt{2\omega_{\bm{k}}}}\psi(\bm{k})
\exp\!\left[\mathrm{i}(\bm{k}\cdot\bm{x}
  -\omega_{\bm{k}}t)\right].
\end{eqnarray}

We note that $\psi(\bm{x},t)$ and $\phi(\bm{x},t)$ coincide in the nonrelativistic limit
in the sense that
\begin{equation}\label{eq:nonrelphi}
  \phi(\bm{x},t)\to\sqrt{\hbar/2mc^2}\psi(\bm{x},t)
\end{equation}
if the $\bm{k}$-space wave packet $\psi(\bm{k})$ is dominated by
low-energy modes $\hbar k\ll mc$.
However, for general $\psi(\bm{k})$, $\psi(\bm{x},t)$ and $\phi(\bm{x},t)$ provide two
different spacetime descriptions of the wave packet $\psi(\bm{k})$. As such, they also
provide two different sets of local densities for energy, momentum, and charge of the wave packet.
The different densities yield the same values for energy, momentum and charge, but
coincide \textit{locally} only in the nonrelativistic limit (\ref{eq:nonrelphi}) or if the
wave packet $\psi(\bm{k})$ is concentrated around a
momentum $\hbar\bm{k}_0$ with small width $|\Delta\bm{k}|\ll|\bm{k}_0|$,
\begin{equation}\label{eq:conck0}
  \phi(\bm{x},t)\to\psi(\bm{x},t)/\sqrt{2\omega(\bm{k}_0)}.
\end{equation}

Before entering the discussion of the different spacetime densities, we note that we can construct
the first quantized field (\ref{eq:phic1}) also through the coherent state
\begin{eqnarray}\nonumber
\bm{|}\psi_c(t)\bm{\rangle}&=&\exp\!\left(\int\!d^3\bm{k}\,a^+(\bm{k})\psi(\bm{k})
\exp(-\,\mathrm{i}\omega_{\bm{k}}t)-\frac{1}{2}\right)\!\bm{|}0\bm{\rangle}
\\ \label{eq:phicoh1}
&=&
\exp\!\left(\int\!d^3\bm{x}\,a^+(\bm{x})\psi(\bm{x},t)-\frac{1}{2}\right)\!\bm{|}0\bm{\rangle},
\end{eqnarray}
through the expectation value of the quantum field in the Dirac interaction picture
or the Schr\"odinger picture, 
\begin{eqnarray}\label{eq:phic2}
  \phi(\bm{x},t)&=&\bm{\langle}\psi_c(0)\bm{|}\Phi(\bm{x},t)\bm{|}\psi_c(0)\bm{\rangle}
  =\bm{\langle}\psi_c(t)\bm{|}\Phi(\bm{x},0)\bm{|}\psi_c(t)\bm{\rangle}.
\end{eqnarray}
The single-particle state $\bm{|}\psi(t)\bm{\rangle}$ (\ref{eq:1p}) is an
eigenstate of the number operator,
\begin{equation}
  \int\!d^3\bm{k}\,a^+(\bm{k})a(\bm{k})\bm{|}\psi(t)\bm{\rangle}=\bm{|}\psi(t)\bm{\rangle},
  \end{equation}
whereas the corresponding coherent state $\bm{|}\psi_c(t)\bm{\rangle}$
(\ref{eq:phicoh1}) (which is normalized due to (\ref{eq:normphi1}))
only satisfies
\begin{equation}
  \bm{\langle}\psi_c(t)\bm{|}\int\!d^3\bm{k}\,a^+(\bm{k})a(\bm{k})\bm{|}\psi_c(t)\bm{\rangle}
  =1.
\end{equation}

The field $\Phi(x)$ in (\ref{eq:KGsolve}) is the quantum field in the Dirac interaction
picture, whereas the quantum field in the Heisenberg picture 
would satisfy the equation
\begin{equation}\label{eq:fullKGH}
\left(\partial-\mathrm{i}\frac{q}{\hbar}A\right)^2\!\Phi_H-\frac{m^2c^4}{\hbar}\Phi_H
-\hbar c^3\frac{\lambda}{2}\Phi_H^+\Phi_H^2=0.
\end{equation}
However, we need to keep in mind that (\ref{eq:fullKGH}) is a nonlinear evolution
equation for the \textit{quantum field} $\Phi_H(x)$ in the Heisenberg picture, but we
\textit{cannot infer} a corresponding nonlinear wave equation for 
a corresponding interacting ``first-quantized'' Klein-Gordon field $\phi_H(x)$.
This is simply a consequence of the fact that
\begin{equation}
\bm{\langle}\Phi_H^+\Phi_H^2\bm{\rangle}\neq\phi_H^+\phi_H^2.
\end{equation}
We can use the nonlinear evolution equation (\ref{eq:fullKGH}) to derive the
Dyson time evolution operator and the scattering matrix of the interacting theory,
but we cannot describe particle interactions through nonlinear evolution
equations for \textit{wave functions}.

\section{Spacetime densities for energy and momentum\label{sec:densities}}

The Hamitonian densities of the Klein-Gordon field in the interaction picture
are $\mathcal{H}(\bm{x},t)$ and  $\mathcal{H}_I(\bm{x},t)$, where the operator
$\mathcal{H}_I(\bm{x},t)$ (\ref{eq:HI}) acts on the states, whereas the evolution
of the interaction picture quantum fields
$\Phi(\bm{x},t)=\exp(\mathrm{i}Ht/\hbar)\Phi(\bm{x},0)\exp(-\,\mathrm{i}Ht/\hbar)$
is governed by the free Hamilton operator\footnote{
  We use the designations $\mathcal{H}$ and $H$ for the Hamiltonian on
  the quantum fields in the interaction picture,
  instead of the usual designations $\mathcal{H}_0$ and $H_0$,
  because it is the evolution of the Hamiltonian density $\mathcal{H}$
  as a measure for signal location, that we want to compare
  with the Born density $|\psi|^2$ and the corresponding energy
  peudo-density $\tilde{\mathcal{H}}$.}
 $H$ with density
\begin{eqnarray} \nonumber
  \mathcal{H}_{\Phi}(\bm{x},t)&=&\frac{m^2c^4}{\hbar}\left|\Phi(\bm{x},t)\right|^2
  +\hbar\dot{\Phi}^+(\bm{x},t)\cdot\dot{\Phi}(\bm{x},t)
\\ \label{eq:KGHdensity}
&&
+\hbar c^2\bm{\nabla}\Phi^+(\bm{x},t)\cdot\bm{\nabla}\Phi(\bm{x},t),
\end{eqnarray}
\begin{eqnarray}
  H&=&\int\!d^3\bm{x}\,\mathcal{H}_{\Phi}(\bm{x},t)
  =\int\!d^3\bm{k}\,\hbar\omega_{\bm{k}}\!\left[
  a^+(\bm{k})a(\bm{k})+b^+(\bm{k})b(\bm{k})\right]\!.
\end{eqnarray}
The momentum density and operator of the interaction picture fields are
\begin{eqnarray}
\bm{\mathcal{P}}_{\Phi}(\bm{x},t)&=&
-\,\hbar\dot{\Phi}^+(\bm{x},t)\cdot\bm{\nabla}\Phi(\bm{x},t)
-\hbar\bm{\nabla}\Phi^+(\bm{x},t)\cdot\dot{\Phi}(\bm{x},t),
\end{eqnarray}
\begin{eqnarray}
  \bm{P}&=&\int\!d^3\bm{x}\,\bm{\mathcal{P}}_{\Phi}(\bm{x},t)
  =\int\!d^3\bm{k}\,\hbar\bm{k}\!\left[
a^+(\bm{k})a(\bm{k})+b^+(\bm{k})b(\bm{k})
\right]\!,
\end{eqnarray}
and the charge density and operator are
\begin{eqnarray}
  \varrho_{\Phi}(\bm{x},t)&=&\mathrm{i}q\left[\Phi^+(\bm{x},t)\cdot\dot{\Phi}(\bm{x},t)
   -\dot{\Phi}^+(\bm{x},t)\cdot\Phi(\bm{x},t)\right]\!,
\end{eqnarray}
\begin{eqnarray} \label{eq:KGcharge}
Q&=&\int\!d^3\bm{x}\,\varrho_{\Phi}(\bm{x},t)
=q\int\!d^3\bm{k}\left[
a^+(\bm{k})a(\bm{k})-b^+(\bm{k})b(\bm{k})\right]\!.
\end{eqnarray}

The energy expectation values $\bm{\langle}E\bm{\rangle}$
(denoted by $E$ for short) both of the single-particle
state (\ref{eq:1p}) and of the coherent state (\ref{eq:phicoh1}) are
\begin{eqnarray}\nonumber
  E&=&\bm{\langle}\psi(t)\bm{|}H\bm{|}\psi(t)\bm{\rangle}
  =\bm{\langle}\psi_c(t)\bm{|}H\bm{|}\psi_c(t)\bm{\rangle}
  \\ \nonumber
  &=&\int\!d^3\bm{x}\,\frac{\mathrm{i}\hbar}{2}\left[
    \psi^+(\bm{x},t)\cdot\dot{\psi}(\bm{x},t)
    -\dot{\psi}^+(\bm{x},t)\cdot
    \psi(\bm{x},t)\right]
  \\ 
  &=&\int\!d^3\bm{k}\,\hbar\omega_{\bm{k}}\left|\psi(\bm{k})\right|^2
  =\int\!d^3\bm{x}\,\mathcal{H}(\bm{x},t),
\end{eqnarray}
with
\begin{eqnarray}\nonumber
  \mathcal{H}(\bm{x},t)&=&\frac{m^2c^4}{\hbar}\left|\phi(\bm{x},t)\right|^2
  +\hbar\dot{\phi}^+(\bm{x},t)\cdot\dot{\phi}(\bm{x},t)
  \\
  &&+\hbar c^2\bm{\nabla}\phi^+(\bm{x},t)\cdot\bm{\nabla}\phi(\bm{x},t).
\end{eqnarray}

The momentum expectation values both of the single-particle
state (\ref{eq:1p}) and of the coherent state (\ref{eq:phicoh1}) are
\begin{eqnarray}\nonumber
  &&\bm{\langle}\psi(t)\bm{|}\bm{P}\bm{|}\psi(t)\bm{\rangle}
  =\bm{\langle}\psi_c(t)\bm{|}\bm{P}\bm{|}\psi_c(t)\bm{\rangle}
  =\int\!d^3\bm{k}\,\hbar\bm{k}\left|\psi(\bm{k})\right|^2
  \\ \nonumber
  &&=\int\!d^3\bm{x}\,\frac{\hbar}{2\mathrm{i}}\left[
    {\psi}^+(\bm{x},t)\cdot\bm{\nabla}
    \psi(\bm{x},t)
    -\bm{\nabla}\psi^+(\bm{x},t)\cdot{\psi}(\bm{x},t)
    \right]
  \\ \label{eq:2rhoP}
  &&=-\,\hbar\int\!d^3\bm{x}\left[
    \dot{\phi}^+(\bm{x},t)\cdot\bm{\nabla}\phi(\bm{x},t)
    +\bm{\nabla}\phi^+(\bm{x},t)\cdot\dot{\phi}(\bm{x},t)
\right]\!.
\end{eqnarray}

The charge expectation values both of the single-particle
state (\ref{eq:1p}) and of the coherent state (\ref{eq:phicoh1}) are
\begin{eqnarray}\nonumber
  \bm{\langle}\psi(t)\bm{|}Q\bm{|}\psi(t)\bm{\rangle}
  &=&\bm{\langle}\psi_c(t)\bm{|}Q\bm{|}\psi_c(t)\bm{\rangle}
  =q\int\!d^3\bm{k}\left|\psi(\bm{k})\right|^2
  \\
  &=&q\int\!d^3\bm{x}\left|\psi(\bm{x},t)\right|^2
  =q\int\!d^3\bm{x}\,\varrho(\bm{x},t),
\end{eqnarray}
where
\begin{eqnarray}
  \varrho(\bm{x},t)&=&\mathrm{i}q\left[\phi^+(\bm{x},t)\cdot\dot{\phi}(\bm{x},t)
   -\dot{\phi}^+(\bm{x},t)\cdot\phi(\bm{x},t)\right]\!.
\end{eqnarray}

In terms of normalization and energy-momentum expectation values, $\psi(\bm{k})$
and $\psi(\bm{x},t)$ are single-particle wave packets in momentum
and position space with energy-momentum densities
\begin{eqnarray} \label{eq:pseudodensity}
  \wp_\mu(\bm{x},t)&=&\frac{\hbar}{2\mathrm{i}}\left(
  \psi^+(\bm{x},t)\cdot\partial_\mu\psi(\bm{x},t)
  -\partial_\mu\psi^+(\bm{x},t)\cdot\psi(\bm{x},t)\right),
\end{eqnarray}
which resemble a relativistic generalization of the nonrelativistic
momentum density (\ref{eq:mathcalP}).

The expectation values lend themselves to the identification of four possible proxies for
position of a signal due to absorption or scattering of a scalar particle, \textit{viz.}
$\mathcal{H}(\bm{x},t)/E$, $c\wp^0(\bm{x},t)/E$, $\varrho(\bm{x},t)/q$
and $|\psi(\bm{x},t)|^2$.

Indeed, we have coincidence in the limits (\ref{eq:nonrelphi})
and (\ref{eq:conck0}),
$\mathcal{H}(\bm{x},t)/E\to c\wp^0(\bm{x},t)/E\to\varrho(\bm{x},t)/q\to |\psi(\bm{x},t)|^2$,
where $E=mc^2$ in the limit (\ref{eq:nonrelphi}) and $E=\hbar\omega(\bm{k}_0)$
in the limit(\ref{eq:conck0}).

If we also have an antiparticle with $\bm{k}$-space
wave function $\chi(\bm{k})$ and Fourier transform
\begin{equation}\label{eq:chixt}
\chi(\bm{x},t)=\frac{1}{\sqrt{2\pi}^3}\int\!d^3\bm{k}\,\chi(\bm{k})
\exp\!\left[\mathrm{i}(\bm{k}\cdot\bm{x}
-\omega_{\bm{k}}t)\right],
\end{equation}
the first-quantized Klein--Gordon field (\ref{eq:phic1}) becomes
\begin{eqnarray}\nonumber
\phi(x)&=&\frac{1}{\sqrt{2\pi}^3}
\int\frac{d^3\bm{k}}{\sqrt{2\omega_{\bm{k}}}}\Big(\psi(\bm{k})
\exp\!\left[\mathrm{i}(\bm{k}\cdot\bm{x}
  -\omega_{\bm{k}}t)\right]
\\ \label{eq:phic12}
&&+\chi^+(\bm{k})
\exp\!\left[-\,\mathrm{i}(\bm{k}\cdot\bm{x}
  -\omega_{\bm{k}}t)\right]\Big),
\end{eqnarray}
the energy of the asymptotic 2-particle state is
\begin{eqnarray}\nonumber
  E&=&\int\!d^3\bm{x}\,\Big[
\hbar\dot{\phi}^+(\bm{x},t)\cdot\dot{\phi}(\bm{x},t)
+\hbar c^2\bm{\nabla}\phi^+(\bm{x},t)\cdot\bm{\nabla}\phi(\bm{x},t)
\\ \nonumber
&&\,\quad+\,\frac{m^2c^4}{\hbar}\left|\phi(\bm{x},t)\right|^2
\Big]
  \\ \nonumber
  &=&\int\!d^3\bm{k}\,\hbar\omega_{\bm{k}}\left(|\psi(\bm{k})|^2+|\chi(\bm{k})|^2\right)
  \\ \nonumber
  &=&\int\!d^3\bm{x}\,\frac{\mathrm{i}\hbar}{2}\big[
    \psi^+(\bm{x},t)\cdot\dot{\psi}(\bm{x},t)
    -\dot{\psi}^+(\bm{x},t)\cdot\psi(\bm{x},t)
  \\  \label{eq:new2rhoH}
  &&+
    \chi^+(\bm{x},t)\cdot\dot{\chi}(\bm{x},t)
    -\dot{\chi}^+(\bm{x},t)\cdot
    \chi(\bm{x},t)\big],
\end{eqnarray}
the momentum is
\begin{eqnarray}\nonumber
  \bm{p}&=&-\,\hbar\!\int\!d^3\bm{x}\left[
    \dot{\phi}^+(\bm{x},t)\cdot\bm{\nabla}\phi(\bm{x},t)
    +\bm{\nabla}\phi^+(\bm{x},t)\cdot\dot{\phi}(\bm{x},t)\right]
  \\ \nonumber
  &=&\int\!d^3\bm{k}\,\hbar\bm{k}\left(|\psi(\bm{k})|^2+|\chi(\bm{k})|^2\right)
  \\ \nonumber
  &=&\int\!d^3\bm{x}\,\frac{\hbar}{2\mathrm{i}}\big[
    {\psi}^+(\bm{x},t)\cdot\bm{\nabla}
    \psi(\bm{x},t)
    -\bm{\nabla}\psi^+(\bm{x},t)\cdot{\psi}(\bm{x},t)
    \\
    &&+ {\chi}^+(\bm{x},t)\cdot\bm{\nabla}
    \chi(\bm{x},t)
    -\bm{\nabla}\chi^+(\bm{x},t)\cdot{\chi}(\bm{x},t)\big],
\end{eqnarray}
and the charge is
\begin{eqnarray}\nonumber
  Q&=&\mathrm{i}q\int\!d^3\bm{x}\left[\phi^+(\bm{x},t)\cdot\dot{\phi}(\bm{x},t)
    -\dot{\phi}^+(\bm{x},t)\cdot\phi(\bm{x},t)\right]
 \\ \nonumber
  &=&q\int\!d^3\bm{k}\left(|\psi(\bm{k})|^2-|\chi(\bm{k})|^2\right)
  \\ 
  &=&q\int\!d^3\bm{x}\left(|\psi(\bm{x},t)|^2-|\chi(\bm{x},t)|^2\right)\!.
\end{eqnarray}

In nonrelativistic quantum mechanics, we accept the very same kind of
relations between the $\bm{x}$-space amplitudes
$\psi(\bm{x},t)$ and $\chi(\bm{x},t)$, and the observables $E$, $\bm{p}$, $Q$,
as evidence for the Born interpretation of 
$\psi(\bm{x},t)$ and $\chi(\bm{x},t)$ as probability amplitudes for
particle position or antiparticle position, respectively.
Why should we not accept this line of reasoning then
also in relativistic quantum theory? 

The Fourier transformed amplitudes $\psi(\bm{x},t)$ and $\chi(\bm{x},t)$
on the one hand, and the Klein--Gordon field $\phi(\bm{x},t)$ on the other hand,
provide different local spacetime expressions for the energy, momentum, and
charge densities of particles.
Substituting the inversion of (\ref{eq:varphixt})
into (\ref{eq:phic1}) shows that the two different kinds of
spacetime fields, \textit{viz.} $\psi(\bm{x},t)$ and $\phi(\bm{x},t)$,
are nonlocally related if the wave packet $\psi(\bm{k})$ is such that
neither of the limits (\ref{eq:nonrelphi}) or (\ref{eq:conck0}) applies.
The question therefore arises how the different
local densities behave for single-particle
wave packets $\psi(\bm{k})$ if the limits (\ref{eq:nonrelphi})
and (\ref{eq:conck0}) do not apply. Formally, we can substitute
the inversion of (\ref{eq:varphixt}) into the energy-momentum and charge densities of
the Klein-Gordon field, but the resulting nonlocal expressions are unwieldy
and do not directly relate the different local expressions for energy-momentum
and charge densities for the single-particle solution.
Therefore, in Sec.~\ref{sec:scalar2} we evaluate
expressions for the different densities numerically.

\section{Comparison of the different energy, momentum and charge
  densities for the massless Klein-Gordon field\label{sec:scalar2}}

We are interested in testing and illustrating the most extreme case
of relativistic wave packet evaluation in the simplest possible setup.
Therefore we assume that our $k$-space wave packet at time $t=0$
is a \textit{massless} scalar meson wave packet
of width $\Delta k=1/2\Delta x$ in one spatial dimension,
\begin{equation}\label{eq:Gaussk0}
  \psi(k,0)=\left(\frac{2\Delta x^2}{\pi}\right)^{1/4}
  \exp(-\,\Delta x^2\cdot k^2).
\end{equation}
The advantage of using the one-dimensional Klein--Gordon field consists
in simpler analytic formulae while exhibiting the same qualitative features
of the different proxies for position that we also find in three spatial
dimensions. These features are easier to illustrate in the one-dimensional
case, because evolution of the packet (\ref{eq:Gaussk0}) yields
position proxies which propagate on or near the light cone.
The features propagating along or near the light cone are more prominent
relative to the maxima of the position proxies at $x=0$, $t=0$ in the
one-dimensional case than in the three-dimensional case, thus yielding
better visibility.

The parameter $\Delta x=1/2\Delta k$ as such is only a placeholder for
momentum uncertainty in the wave packet (\ref{eq:Gaussk0}). $\Delta x$ corresponds
to a \textit{proxy} for position uncertainty if we use the square $|\psi(x,0)|^2$
of the Fourier transformed wave packet $\psi(x,0)$ as a proxy for position
at $t=0$.
However, we cannot address $\Delta x$ as an actual position uncertainty
because we cannot address $|\psi(x,t)|^2$ as a proper probability density for position.

Eq.~(\ref{eq:Gaussk0})  corresponds to a Gaussian superposition of massless meson
normal modes with an energy
\begin{equation}\label{eq:EGaussk}
  E=\int\!dk\,\hbar c|k|\,|\psi(k,t)|^2=\frac{\hbar c}{\sqrt{2\pi\Delta x^2}}.
\end{equation}

Eq.~(\ref{eq:Gaussk0}) also implies that the Fourier transformed wave packet
at $t=0$ is a Gaussian wave packet \textit{at rest},
\begin{equation}\label{eq:Gaussx0}
\psi(x,0)=\frac{\exp(-\,x^2/4\Delta x^2)}{(2\pi\Delta x^2)^{1/4}}.
\end{equation}
The initial condition (\ref{eq:Gaussk0}) 
may therefore appear self-contradictory from the outset, because
massless particles always move at the speed of light.
However, the free Klein--Gordon equation can evolve \textit{any}
initial wave packet, even in the massless limit, and it must
clearly also be able to consistently evolve the wave
packet (\ref{eq:Gaussx0}) as a superposition of quanta with
2-momenta $(\omega_k/c,k)=(|k|,k)$.
Indeed, the massless limit of the relativistic dispersion relation
is
\begin{equation}
\omega_k=\lim_{m\to 0}c\sqrt{k^2+(mc/\hbar)^2}=c|k|,
\end{equation}
and therefore our $k$-space meson wave function at arbitrary time $t$ is
\begin{equation}\label{eq:Gaussk}
  \psi(k,t)=\left(\frac{2\Delta x^2}{\pi}\right)^{1/4}
  \exp(-\,\Delta x^2 k^2-\mathrm{i}c|k|t).
\end{equation}
The Fourier transformed wave packet at time $t$ involves the complex error function
\begin{equation}
  \mathrm{erfi}(z)=-\,\mathrm{i}\cdot\mathrm{erf}(\mathrm{i}z)
  =\frac{2}{\mathrm{i}\sqrt{\pi}}\int_0^{\mathrm{i}z}\!du\,\exp(-\,u^2),
\end{equation}
such that
\begin{eqnarray}\nonumber
  \psi(x,t)&=&\frac{1}{2(2\pi\Delta x^2)^{1/4}}\exp\!\left(-\,\frac{(x-ct)^2}{4\Delta x^2}\right)
  \left[1+\mathrm{i}\cdot\mathrm{erfi}\!\left(\frac{x-ct}{2\Delta x}\right)\right]
      \\ \label{eq:Psi0xt}
      &&+\frac{1}{2(2\pi\Delta x^2)^{1/4}}\exp\!\left(-\,\frac{(x+ct)^2}{4\Delta x^2}\right)
      \left[1-\mathrm{i}\cdot\mathrm{erfi}\!\left(\frac{x+ct}{2\Delta x}\right)\right]\!.
\end{eqnarray}
On the other hand, the Klein-Gordon field with the amplitude (\ref{eq:Gaussk}),
\begin{eqnarray} \label{eq:PhiGauss}
  \phi(x,t)&=&\left(\frac{\Delta x^2}{8\pi^3 c^2}\right)^{1/4}
  \int_{-\infty}^{\infty}\!\frac{dk}{\sqrt{|k|}}
  \exp(\mathrm{i}kx-\Delta x^2 k^2-\mathrm{i}c|k|t),
\end{eqnarray}
involves modified Bessel functions of the first kind,
\begin{eqnarray}\nonumber
  \phi(x,t)&=&\frac{1}{4}\left(\frac{\pi}{2c^2\Delta x^2}\right)^{1/4}
  \exp\!\left(-\,\frac{(x-ct)^2}{8\Delta x^2}\right)
  \\ \nonumber
  &&\times\!\left[\sqrt{|x-ct|}I_{-\frac{1}{4}}\!\left(\frac{(x-ct)^2}{8\Delta x^2}\right)\!
    +\mathrm{i}
   \frac{x-ct}{\sqrt{|x-ct|}}I_{\frac{1}{4}}\!\left(\frac{(x-ct)^2}{8\Delta x^2}\right)
    \right]
  \\ \nonumber
  &&+\frac{1}{4}\left(\frac{\pi}{2c^2\Delta x^2}\right)^{1/4}
  \exp\!\left(-\,\frac{(x+ct)^2}{8\Delta x^2}\right)
  \\ \label{eq:Phi0xt}
  &&\times\!\left[\sqrt{|x+ct|}I_{-\frac{1}{4}}\!\left(\frac{(x+ct)^2}{8\Delta x^2}\right)\!
   -\mathrm{i}
   \frac{x+ct}{\sqrt{|x+ct|}}I_{\frac{1}{4}}\!\left(\frac{(x+ct)^2}{8\Delta x^2}\right)
    \right]\!.
\end{eqnarray}


The asymptotic behavior of the normalized wave packet $\psi(x,t)$ (\ref{eq:Psi0xt})
for $|x|\gg |ct|$ is proportional to $\exp(-\,x^2/4\Delta x^2)$, whereas
the asymptotic behavior of the Klein--Gordon wave packet (\ref{eq:Phi0xt})
is given by $|\phi(x,t)|\propto\sqrt{\Delta x/c|x|}$, whence
$\int_{-x}^x dy\,|\phi(y,t)|^2$ diverges logarithmically.
Qualitatively, the
suppression of the high frequency modes with $1/\sqrt{|k|}$ in $\phi(x,t)$
versus $\psi(x,t)$ implies less wavy behavior in $\phi(x,t)$ and therefore
less destructive interference for large $|x|$, hence a smaller rate of decrease
for $|x|\to\infty$. 

The $x$-space wave packet $\psi(x,t)$ (\ref{eq:Psi0xt}) and the corresponding
Klein--Gordon field $\phi(x,t)$ (\ref{eq:Phi0xt}) are nonlocally related through
(\ref{eq:Gaussk}). The question therefore arises: What do the pseudo-probability
density $|\psi(x,t)|^2$, the normalized charge density of the Klein--Gordon field,
\begin{eqnarray}\label{eq:normrho}
  \varrho({x},t)/q&=&\mathrm{i}\left[\phi^+({x},t)\dot{\phi}({x},t)
    -\dot{\phi}^+({x},t)\phi({x},t)\right]\!,
\end{eqnarray}
the energy density of the Klein--Gordon field,
\begin{eqnarray}\label{eq:canH}
\mathcal{H}({x},t)&=&
\hbar\dot{\phi}^+({x},t)\dot{\phi}({x},t)
+\hbar c^2\phi'^+({x},t)\phi'({x},t),
\end{eqnarray}
and the energy
pseudo-density $\tilde{\mathcal{H}}({x},t)=c\wp^0(x,t)$ (\ref{eq:pseudodensity})
of the $x$-space wave packet (\ref{eq:Psi0xt}),
\begin{eqnarray}\label{eq:tildeH}
  \tilde{\mathcal{H}}({x},t)&=&\frac{\mathrm{i}\hbar}{2}
  \left[\psi^+({x},t)\dot{\psi}({x},t)
    -\dot{\psi}^+({x},t)\psi({x},t)\right]\!,
\end{eqnarray}
tell us about meson location?

Recall that
\begin{equation}
\int\!dx\,|\psi(x,t)|^2=\int\!dx\,\varrho({x},t)/q=1
\end{equation}
and
\begin{equation}
  \int\!dx\,\mathcal{H}({x},t)=\int\!dx\,\tilde{\mathcal{H}}({x},t)=E,
\end{equation}
where $E$ is given by (\ref{eq:EGaussk}).
The four normalized
densities $|\psi(x,t)|^2$, $\varrho({x},t)/q$, $\mathcal{H}({x},t)/E$,
$\tilde{\mathcal{H}}({x},t)/E$ for the wave packet (\ref{eq:Gaussk}) are
displayed in the following plots.

\begin{figure}[htb]\begin{center}
\scalebox{1}{\includegraphics{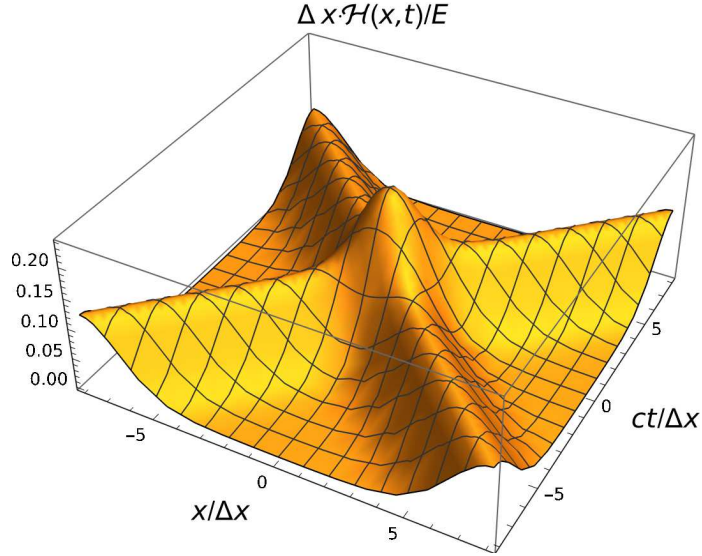}}
\caption{\label{fig:NormHPhixt}
  The normalized canonical energy density $\mathcal{H}(x,t)/E$ (\ref{eq:canH})
  of the massless Klein-Gordon field
  with Gaussian wave packet (\ref{eq:Gaussk}) and energy (\ref{eq:EGaussk}).}
\end{center}\end{figure}

The normalized canonical energy density for the massless Klein--Gordon field
with $k$-space amplitude (\ref{eq:Gaussk}) is shown in Fig.~\ref{fig:NormHPhixt}.
  
\begin{figure}[htb]\begin{center}
\scalebox{1}{\includegraphics{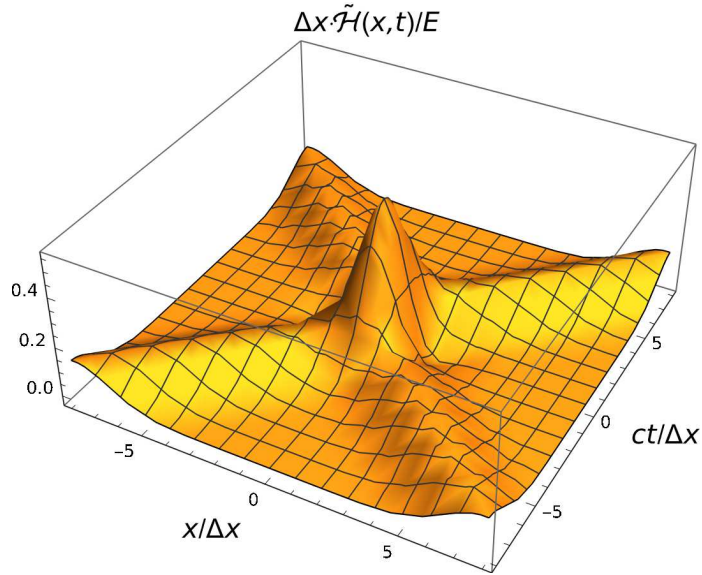}}
\caption{\label{fig:NormTildeHxt}
  The normalized energy pseudo-density $\tilde{\mathcal{H}}(x,t)/E$ (\ref{eq:tildeH})
  of the massless Klein-Gordon field
  with Gaussian wave packet (\ref{eq:Gaussk}). Note that integration of this energy
  pseudo-density for fixed time $t$ also yields the energy (\ref{eq:EGaussk}).}
 \end{center}\end{figure}

The energy pseudo-density from the Fourier transform of the $k$-space
amplitude (\ref{eq:Gaussk}) is shown in Fig.~\ref{fig:NormTildeHxt}

\begin{figure}[htb]\begin{center}
\scalebox{1}{\includegraphics{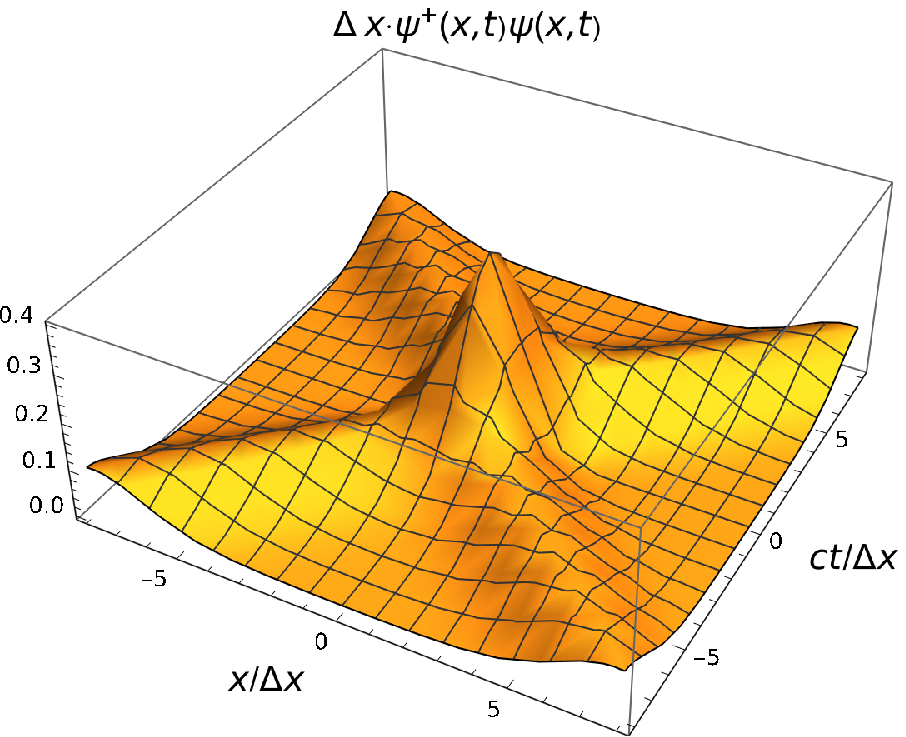}}
\caption{\label{fig:PsiKG0squared2}
  The probability pseudo-density $|\psi(x,t)|^2$ from the Fourier transformed
  $k$-space amplitude of the massless Klein-Gordon field
  with Gaussian wave packet (\ref{eq:Gaussk}).}
  \end{center}\end{figure}

 Furthermore, the probability pseudo-density $|\psi(x,t)|^2$ from the Fourier
 transformation of the $k$-space amplitude (\ref{eq:Gaussk}) is shown
 in Fig.~\ref{fig:PsiKG0squared2}.
 
\begin{figure}[htb]\begin{center}
\scalebox{1}{\includegraphics{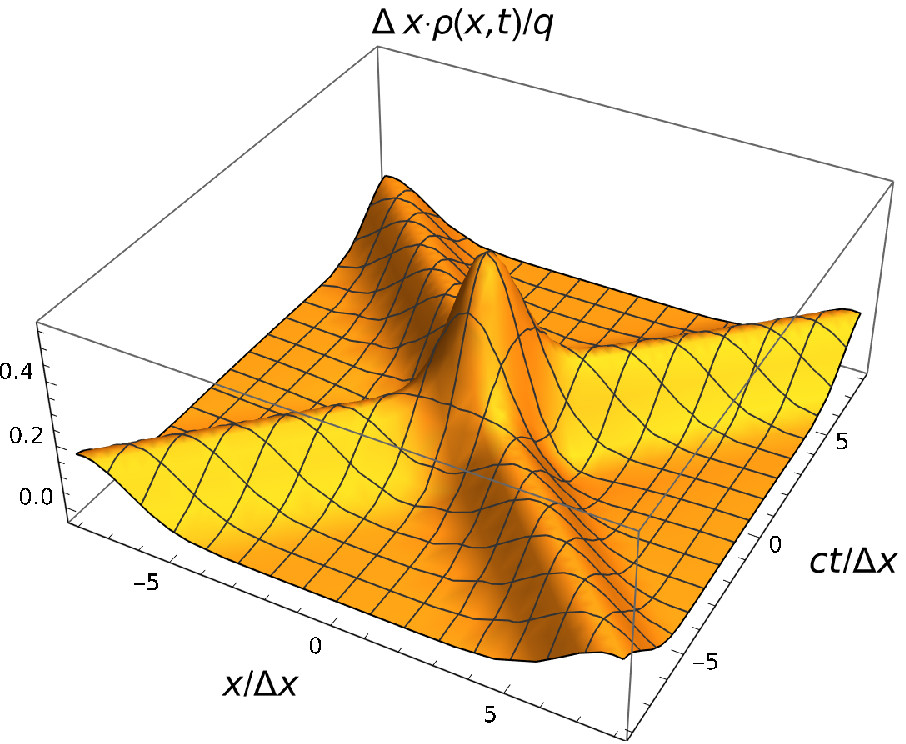}}
\caption{\label{fig:FullRhoxt}
  The normalized charge density $\rho(x,t)/q$ (\ref{eq:normrho})
  for the massless Klein-Gordon field
  with Gaussian wave packet (\ref{eq:Gaussk}).}
  \end{center}\end{figure}

  Finally, Fig.~\ref{fig:FullRhoxt} shows the normalized charge density (\ref{eq:normrho})
  of the Klein--Gordon field with $k$-space amplitude (\ref{eq:Gaussk}).
  
  Although the Fourier tansformed amplitude $\psi(x,t)$ and the
  Klein--Gordon field $\phi(x,t)$, as well as all four considered proxies for
  particle motion, are nonlocally related, they all exhibit the same feature of
  motion along the light cone. This is a consequence of the fact that
  both $\psi(x,t)$ and $\phi(x,t)$ are composed of normal modes $\exp[\mathrm{i}(kx-c|k|t)]$
  that move along the light cone. The physical picture would be that a massless particle
  would be created in the point $x=0$ at $t=0$, and this would
  spread along the light cone similar to a photon that would be created. Of course, if
  instead we would wish to construct the wave packet (\ref{eq:Gaussk0}) at $t=0$ from
  incoming massless normal modes, we would need to superimpose signals coming in along the
  backwards light cone.
  
\begin{figure}[htb]\begin{center}
\scalebox{1}{\includegraphics{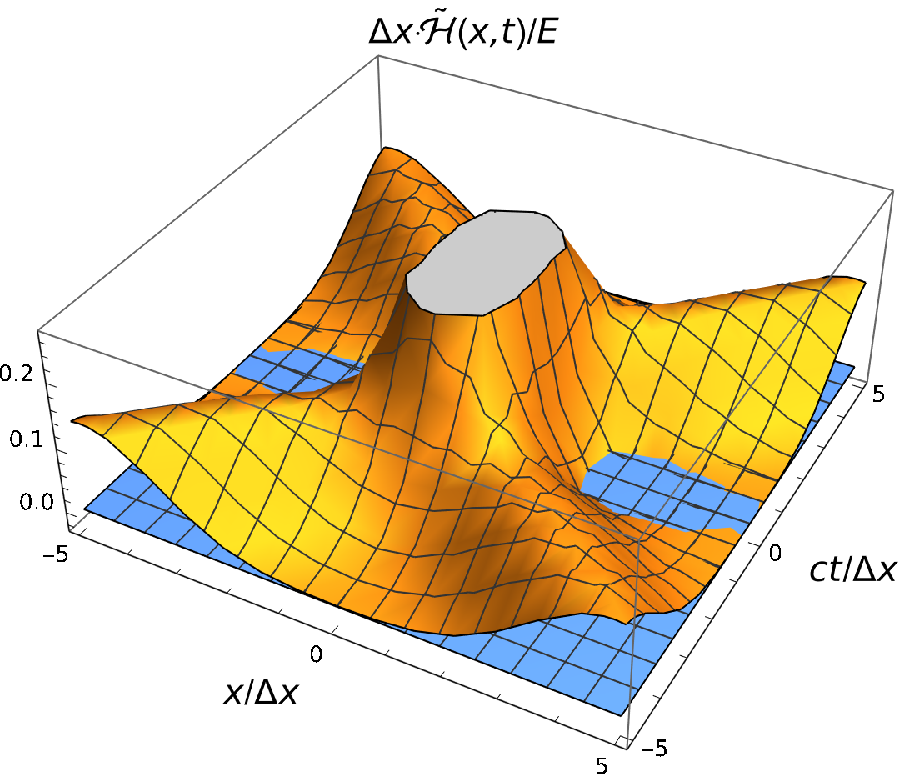}}
\caption{\label{fig:NormTildeH0}
  The normalized energy pseudo-density $\tilde{\mathcal{H}}(x,t)/E$ (\ref{eq:tildeH})
  for the massless Klein-Gordon field with Gaussian wave packet (\ref{eq:Gaussk})
  and energy (\ref{eq:EGaussk}).}
  \end{center}\end{figure}

\begin{figure}[htb]\begin{center}
\scalebox{1}{\includegraphics{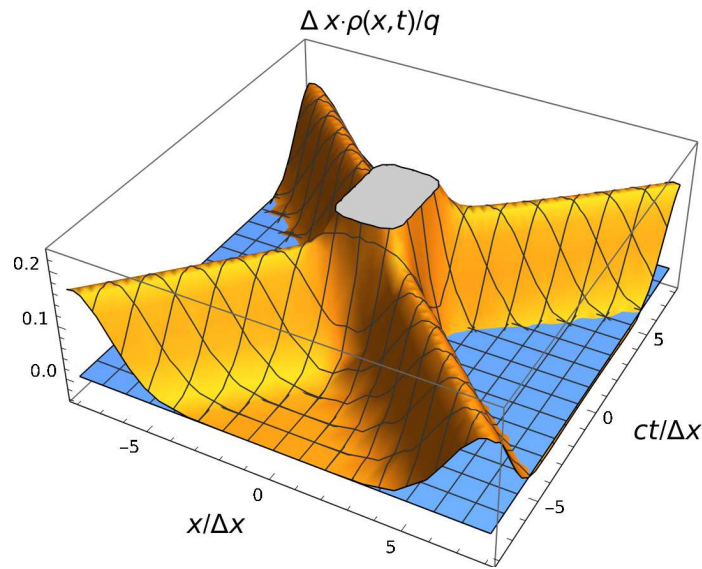}}
\caption{\label{fig:rhoKG0}
  The normalized charge density $\rho(x,t)/q$ (\ref{eq:normrho})
  for the massless Klein-Gordon field
  with Gaussian wave packet (\ref{eq:Gaussk}).}
  \end{center}\end{figure}

  The canonical energy density (\ref{eq:canH}) and the probability pseudo-density
  $|\psi(x,t)|^2$ are manifestly positive semidefinite, but this does not apply
  to the energy pseudo-density (\ref{eq:tildeH}) nor the charge density (\ref{eq:normrho}).
  Naive physical intuition might make us expect that $\tilde{\mathcal{H}}(x,t)$
  and $\varrho(x,t)/q$ should be positive semidefinite nonetheless, because they were
  calculated for the situation where the Klein--Gordon field (\ref{eq:PhiGauss})
  contains only particle contributions, but no antiparticle contributions.
  However, this naive expectation is wrong, as demonstrated in Figs.~\ref{fig:NormTildeH0}
  and \ref{fig:rhoKG0}, where the planes $\tilde{\mathcal{H}}(x,t)=0$
  and $\rho(x,t)=0$ are included in blue. There are regions outside of the light cone
  where $\tilde{\mathcal{H}}(x,t)<0$ or $\rho(x,t)<0$.

\begin{figure}[htb]\begin{center}
\scalebox{1}{\includegraphics{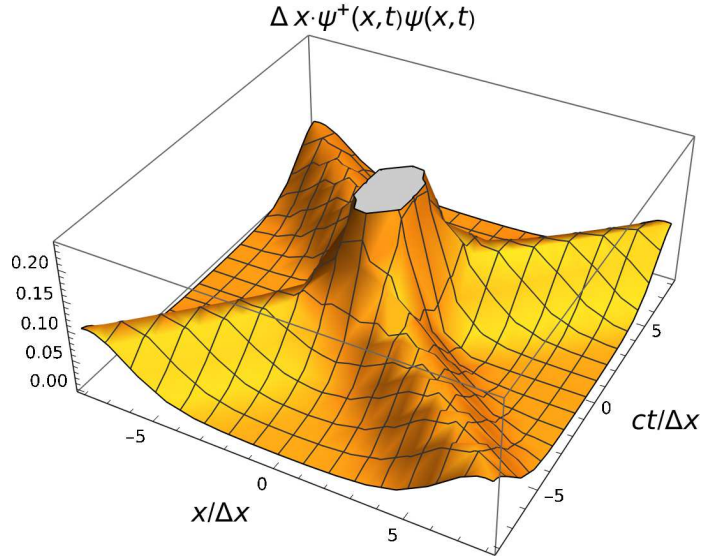}}
\caption{\label{fig:PsiKGsquared3}
  The probability pseudo-density $|\psi(x,t)|^2$ from the Fourier transformed
  $k$-space amplitude of the massless Klein-Gordon field
  with Gaussian wave packet (\ref{eq:Gaussk}).}
  \end{center}\end{figure}

  Therefore only $\mathcal{H}(x,t)/E$ or $|\psi(x,t)|^2$ could possibly serve as fundamental
  probability densities for particle motion. However, just like $\tilde{\mathcal{H}}(x,t)/E$
  and $\rho(x,t)/q$, the probability pseudo-density $|\psi(x,t)|^2$ also exhibits a latency
  effect: The incoming signal, while moving at the speed of light, travels slightly inside of
  the backward light cone of the point $(x,t)=(0,0)$, while the outgoing signal travels inside
  the forward light cone of the point $(x,t)=(0,0)$. This can be inferred from the cut sections
  of the peaks in Figs.~\ref{fig:NormTildeH0}, \ref{fig:rhoKG0}, and \ref{fig:PsiKGsquared3}.
  The offset of the local maxima of those three proxies for particle location from the
  light cone $x^2=c^2t^2$ is always within the proxy $\Delta x$ for position uncertainty
  of the wave packet.
  From the point of view of these three proxies for particle location, the wave packet
  at $x=0$, $t=0$ lingers for an extra time of order $\Delta t\sim\Delta x/c$ before
  it splits for motion along the forward light cone.

\begin{figure}[htb]\begin{center}
\scalebox{1}{\includegraphics{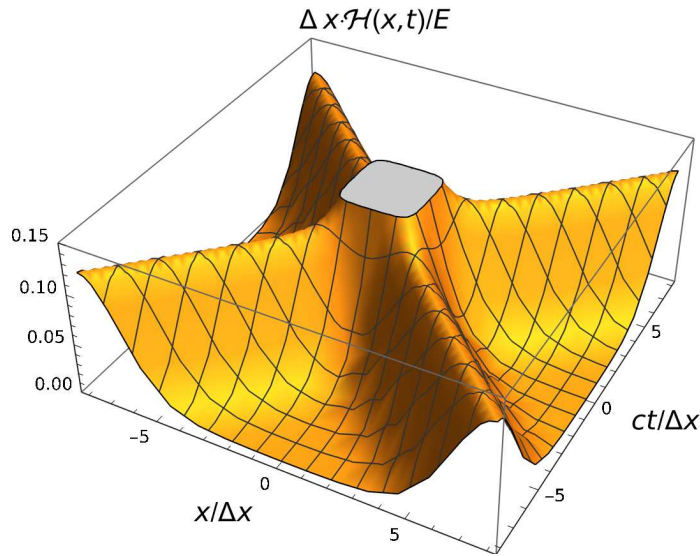}}
\caption{\label{fig:CutNormH}
  The normalized canonical energy density $\mathcal{H}(x,t)/E$ (\ref{eq:canH})
  of the massless Klein-Gordon field
  with Gaussian wave packet (\ref{eq:Gaussk}) and energy (\ref{eq:EGaussk}).}
\end{center}\end{figure}

   On the other hand, the canonical energy density for the massless Klein--Gordon field is
   centered around the light cone, see Fig.~\ref{fig:CutNormH}.

\begin{figure}[htb]\begin{center}
\scalebox{1}{\includegraphics{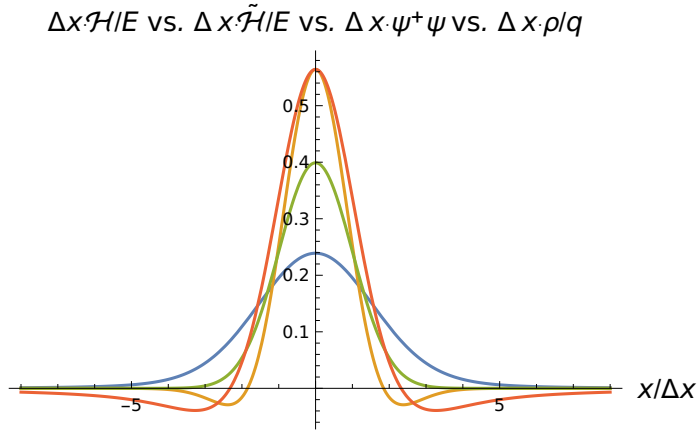}}
\caption{\label{fig:CompAllt0}
  The normalized canonical energy density $\mathcal{H}(x,t)/E$ (\ref{eq:canH})
  (blue), the normalized energy pseudo-density $\tilde{\mathcal{H}}(x,t)/E$ (\ref{eq:tildeH})
  (yellow), the probability pseudo-density $|\psi(x,t)|^2$ (green), and the normalized
  charge density $\rho(x,t)/q$ (\ref{eq:normrho}) (red) at $t=0$.}
\end{center}\end{figure}

\begin{figure}[htb]\begin{center}
\scalebox{1}{\includegraphics{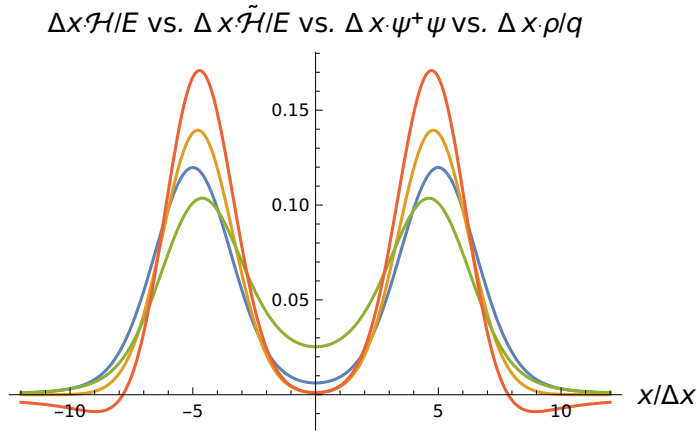}}
\caption{\label{fig:CompAllt5}
  The normalized canonical energy density $\mathcal{H}(x,t)/E$ (\ref{eq:canH})
  (blue), the normalized energy pseudo-density $\tilde{\mathcal{H}}(x,t)/E$ (\ref{eq:tildeH})
  (yellow), the probability pseudo-density $|\psi(x,t)|^2$ (green), and the normalized
  charge density $\rho(x,t)/q$ (\ref{eq:normrho}) (red) at $t=\pm\, 5\Delta x/c$.}
    \end{center}\end{figure}

    Spatial cross sections through all four proxies for particle location are shown for
    $t=0$ in Fig.~\ref{fig:CompAllt0} and for $t=\pm\, 5\Delta x/c$ in
    Fig.~\ref{fig:CompAllt5}. Fig.~\ref{fig:CompAllt5} confirms
    that $\tilde{\mathcal{H}}(x,t)/E$, $|\psi(x,t)|^2$, and $\rho(x,t)/q$ touch
    the light cone from inside within the proxy $\Delta x$ for position uncertainty .
    
    The pattern visible in Figs.~\ref{fig:CompAllt0} and \ref{fig:CompAllt5} 
    persists at all times in that the ``qualitative markings'' of particle position through
    the maxima of any of the position proxies $\mathcal{H}(x,t)/E$, $\tilde{\mathcal{H}}(x,t)/E$,
    $|\psi(x,t)|^2$ and $\rho(x,t)/q$ remain within the uncertainty measure $\Delta x$
    built into the initial wave packet (\ref{eq:Gaussk0}). From
    a practical operational point of view, relativistic quantum mechanics is remarkably robust
    in the sense that any of the four proxies will correctly ``predict'' the signal position
    within the measure $\Delta x$ for position uncertainty,
    even in the massless case and in spite of the complicated
    nonlocal relations between them. Mathematical nonlocality can still yield physical locality.

    Only $\mathcal{H}(x,t)/E$ or $|\psi(x,t)|^2$ could possibly play the role of a fundamental
    probability density for providing predictions for signal locations, instead of only
    providing a proxy. However, they agree within the proxy $\Delta x$ for position
    uncertainty,
    and therefore one might infer that we cannot decide which of those
    two quantities should provide a ``true'' probability density for particle position.
    We will revisit this question in Sec.~\ref{sec:conc}.

\section{Origin of the latency effects\label{sec:latency}}

\begin{figure}[htb]\begin{center}
\scalebox{1}{\includegraphics{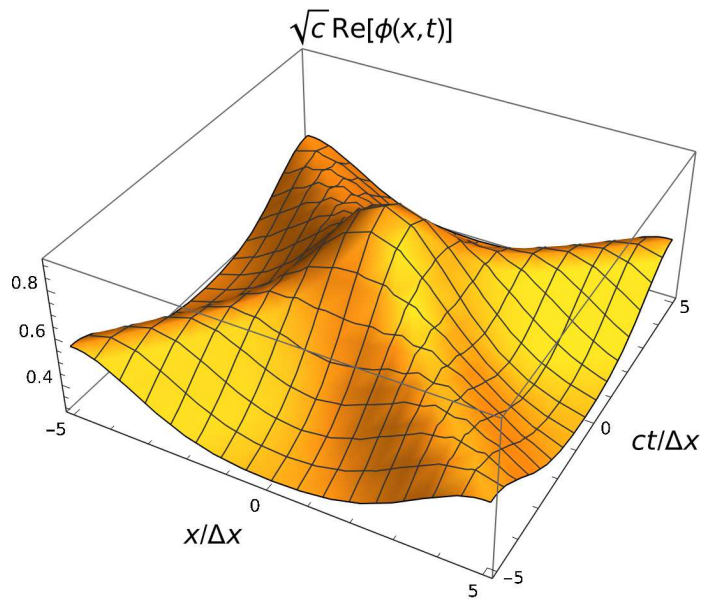}}
\caption{\label{fig:RePhi0}
  The real part of the Klein--Gordon field (\ref{eq:Phi0xt})
  with Gaussian wave packet (\ref{eq:Gaussk}). This part is concentrated along the
  light cone.}
    \end{center}\end{figure}

\begin{figure}[htb]\begin{center}
\scalebox{1}{\includegraphics{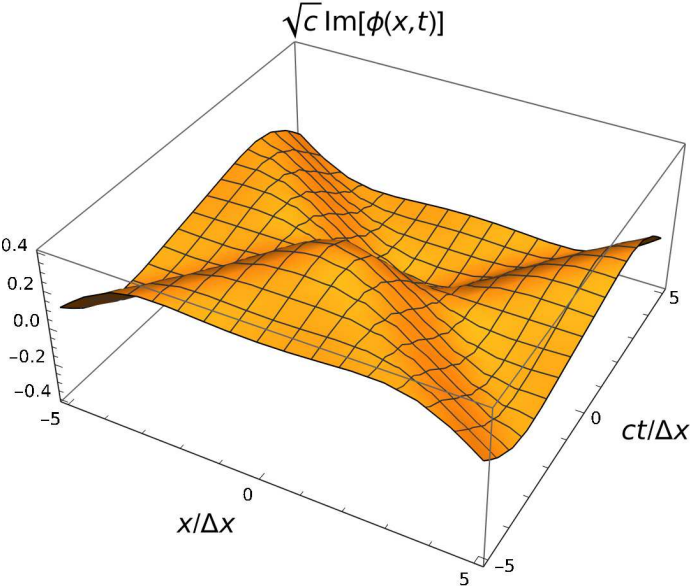}}
\caption{\label{fig:ImPhi0}
  The imaginary part of the Klein--Gordon field (\ref{eq:Phi0xt})
  with Gaussian wave packet (\ref{eq:Gaussk}). This part has the steepest gradient along
  the light cone and larger magnitude inside the light cone.}
    \end{center}\end{figure}

\begin{figure}[htb]\begin{center}
\scalebox{1}{\includegraphics{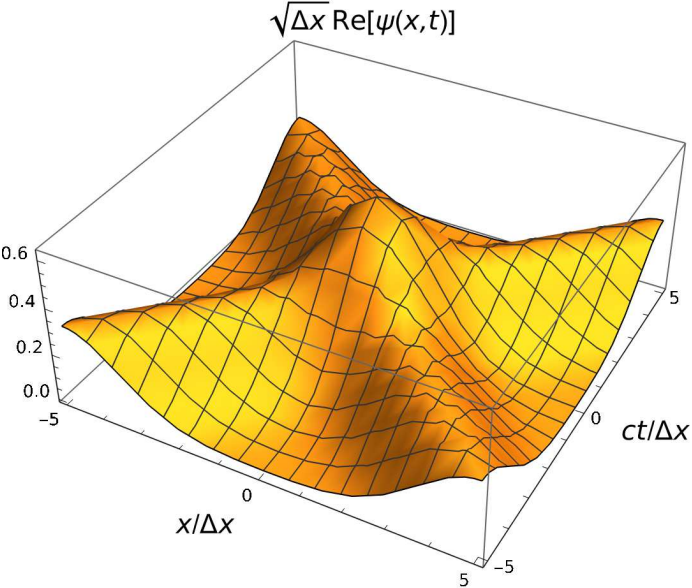}}
\caption{\label{fig:RePsi0}
  The real part of the Fourier transform (\ref{eq:Psi0xt})
  of the Gaussian wave packet (\ref{eq:Gaussk}). This part is concentrated along the
  light cone.}
    \end{center}\end{figure}

\begin{figure}[htb]\begin{center}
\scalebox{1}{\includegraphics{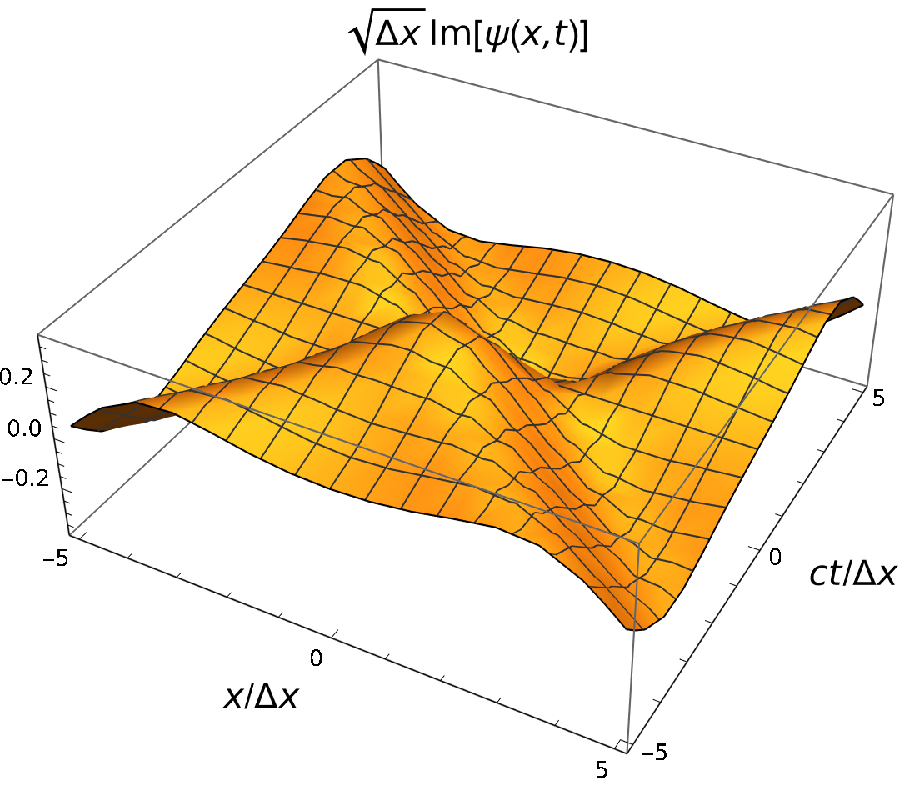}}
\caption{\label{fig:ImPsi0}
  The imaginary part of the Fourier transform (\ref{eq:Psi0xt})
  of the Gaussian wave packet (\ref{eq:Gaussk}). This part has the steepest gradient along
  the light cone and larger magnitude inside the light cone.}
    \end{center}\end{figure}

The latency effects in $|\psi(x,t)|^2$ (Fig.~\ref{fig:PsiKGsquared3})
and in the canonical charge density (\ref{eq:normrho}) (see Fig.~\ref{fig:rhoKG0})
are consequences of the facts that the imaginary contributons both to the
canonical Klein--Gordon field (\ref{eq:Phi0xt}) and to the corresponding wave packet
(\ref{eq:Psi0xt}) are concentrated inside the light cone, whereas the real parts are
concentrated on the light cone.
This is illustrated in Figs.~\ref{fig:RePhi0}--\ref{fig:ImPsi0}.

   Comparison of Fig.~\ref{fig:RePhi0} with Fig.~\ref{fig:RePsi0}, and comparison
   of Fig.~\ref{fig:ImPhi0} with Fig.~\ref{fig:ImPsi0}, shows that $\phi(x,t)$
   and $\psi(x,t)$ have very similar features, but oscillations are more pronounced
   in $\psi(x,t)$. This can be understood as a concequence of the fact that high-frequency
   modes are suppressed in $\phi(x,t)$ relative to $\psi(x,t)$ through the additional factor
   $1/\sqrt{2|k|}$ in Eq.~(\ref{eq:PhiGauss}).
   
   The larger magnitude of $|\mathrm{Im}[\psi(x,t)]|$ inside the light cone immediately explains
   the latency effect in $|\psi(x,t)|^2$ that is visible in Fig.~\ref{fig:PsiKGsquared3}.

   The larger magnitude of $|\mathrm{Im}[\phi(x,t)]|$ inside the light cone implies that
   the term $2\,\mathrm{Im}[\phi(x,t)]\cdot\mathrm{Re}[\dot{\phi}(x,t)]$ in
   \begin{eqnarray}
     \varrho(x,t)/q&=&2\,\mathrm{Im}[\phi(x,t)]\cdot\mathrm{Re}[\dot{\phi}(x,t)]
     -2\,\mathrm{Re}[\phi(x,t)]\cdot\mathrm{Im}[\dot{\phi}(x,t)]
   \end{eqnarray}
   pulls the charge density towards the inside of the light cone. The
   term $-\,2\,\mathrm{Re}[\phi(x,t)]\cdot\mathrm{Im}[\dot{\phi}(x,t)]$
   has maximal magnitude on the light cone.

\section{Effects of mass or momentum\label{sec:mass}}

\begin{figure}[htb]\begin{center}
\scalebox{0.9}{\includegraphics{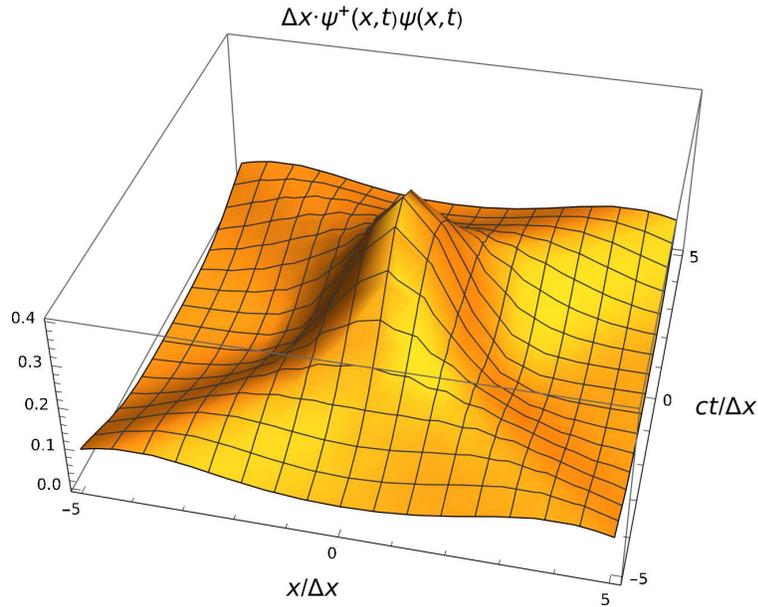}}
\caption{\label{fig:smallmpsisquared}
  The probability pseudo-density $|\psi(x,t)|^2$ from the Fourier transformed
  $k$-space amplitude of the massive Klein-Gordon field ($m=0.2\hbar/c\Delta x$)
  with Gaussian wave packet (\ref{eq:Gausskm}).}
  \end{center}\end{figure}

\begin{figure}[htb]\begin{center}
\scalebox{1}{\includegraphics{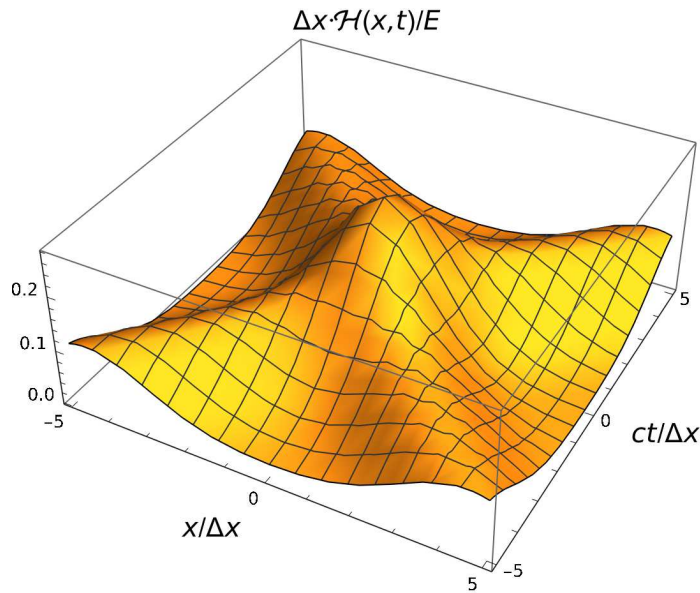}}
\caption{\label{fig:smallmH}
  The normalized canonical energy density $\mathcal{H}(x,t)/E$
  of the massive Klein-Gordon field ($m=0.2\hbar/c\Delta x$)
  with Gaussian wave packet (\ref{eq:Gausskm}).}
  \end{center}\end{figure}

Turning on a mass of the Klein--Gordon field bends the maxima of the propagating features
into the light cone. This is illustrated for
\begin{eqnarray} \label{eq:Gausskm}
  \psi(k,t)&=&\left(\frac{2\Delta x^2}{\pi}\right)^{1/4}\exp\!\left(-\,\Delta x^2 k^2\right)
  \exp\!\left(-\,\mathrm{i}c\sqrt{k^2+(mc/\hbar)^2}t\right),
\end{eqnarray}
with mass $m=0.2\hbar/c\Delta x$,
in Fig.~\ref{fig:smallmpsisquared} for $|\psi(x,t)|^2$ and
in Fig.~\ref{fig:smallmH} for the canonical energy density $\mathcal{H}(x,t)$
of the Klein--Gordon field.

  Figs.~\ref{fig:smallmpsisquared}
  and \ref{fig:smallmH} still show features of the (necessarily ultrarelativistic)
  massless case because $\Delta x=0.2\hbar/mc$ implies for the
  Gaussian wave packet (\ref{eq:Gaussk}) $\Delta k=2.5mc/\hbar$, i.e.~the wave packet
  contains many highly relativistic modes.

\begin{figure}[htb]\begin{center}
\scalebox{1}{\includegraphics{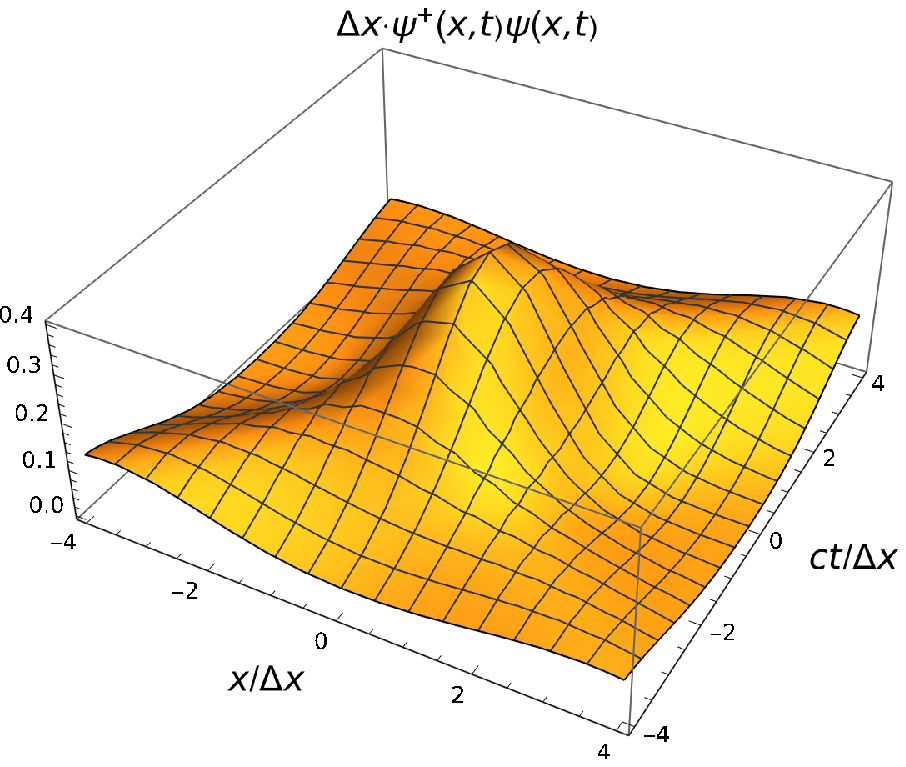}}
\caption{\label{fig:smallK0sisquared}
  The probability pseudo-density $|\psi(x,t)|^2$ from the Fourier transformed
  $k$-space amplitude of the massless Klein-Gordon field
  with Gaussian wave packet (\ref{eq:Gausskk0}) and $k_0=0.1/\Delta x$.}
  \end{center}\end{figure}

\begin{figure}[htb]\begin{center}
\scalebox{1}{\includegraphics{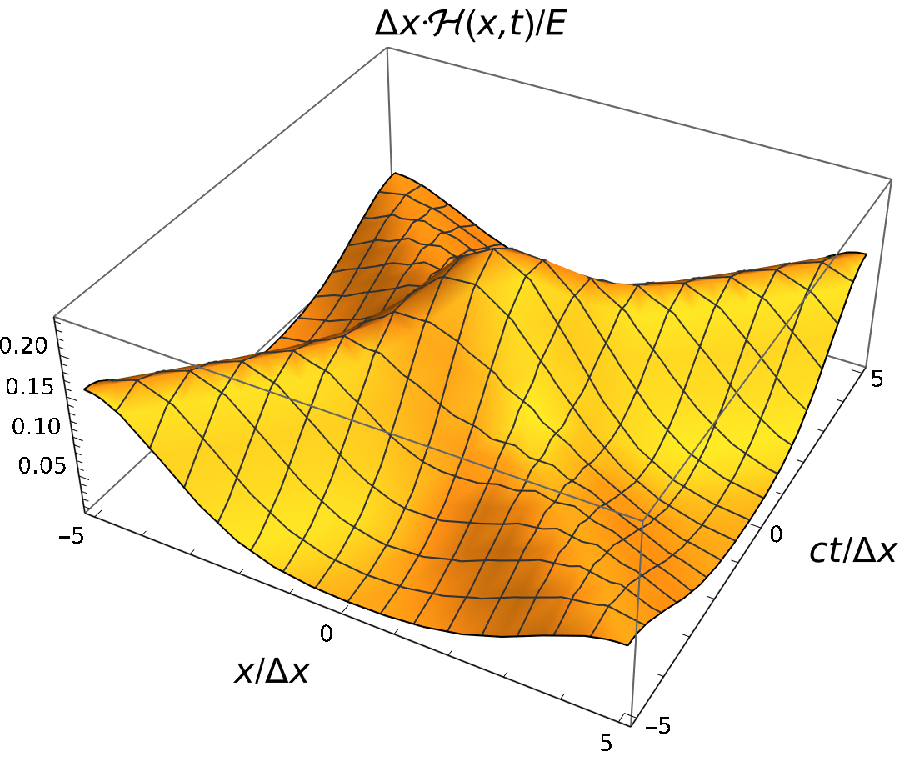}}
\caption{\label{fig:smallK0H}
  The normalized canonical energy density $\mathcal{H}(x,t)/E$
  of the massless Klein-Gordon field 
  with Gaussian wave packet (\ref{eq:Gausskk0}) and $k_0=0.1/\Delta x$.}
  \end{center}\end{figure}

  On the other hand, turning on even only a small center of mass
  momentum $\hbar k_0=0.1\hbar/\Delta x$,
\begin{equation}\label{eq:Gausskk0}
  \psi(k,t)=\left(\frac{2\Delta x^2}{\pi}\right)^{1/4}
  \exp[-\,\Delta x^2 (k-k_0)^2-\mathrm{i}c|k|t],
\end{equation}
  already strongly suppresses the
  component of the wave packet that moves in the opposite direction because the
  negative $k$-components in the wave packet are suppressed relative to the
  positive $k$-components. 
  This is illustrated for $|\psi(x,t)|^2$
  in Fig.~\ref{fig:smallK0sisquared} and for the canonical energy density $\mathcal{H}(x,t)$
  of a massless Klein--Gordon field in Fig.~\ref{fig:smallK0H}.

\begin{figure}[htb]
\begin{center}\scalebox{1}{\includegraphics{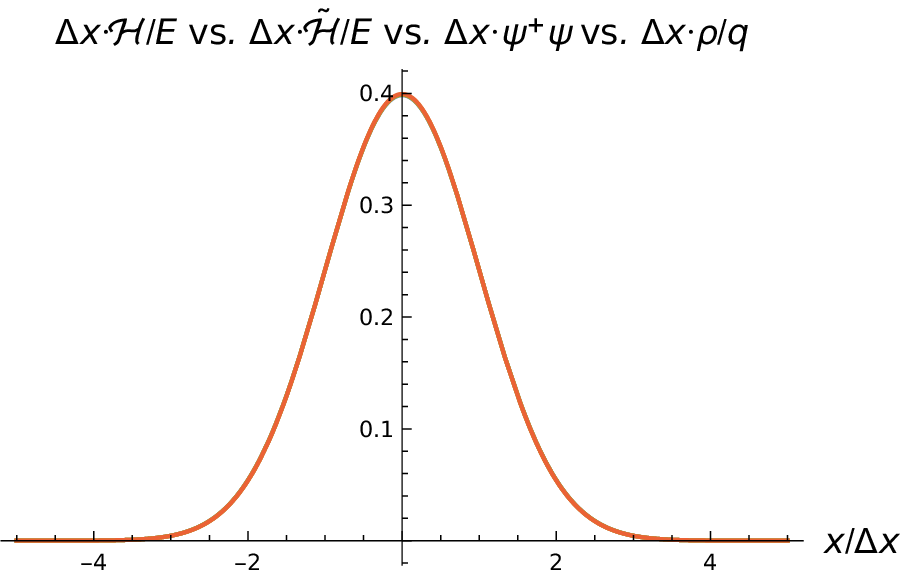}}
\caption{\label{fig:CompAllLargeK}
  The normalized canonical energy density $\mathcal{H}(x,0)/E$ (\ref{eq:canH}),
  the normalized energy pseudo-density $\tilde{\mathcal{H}}(x,0)/E$ (\ref{eq:tildeH}),
  the probability pseudo-density $|\psi(x,0)|^2$, and the normalized
  charge density $\rho(x,0)/q$ (\ref{eq:normrho}) for the massless Klein--Gordon field
   with Gaussian wave packet (\ref{eq:Gausskk0}) and $k_0=10/\Delta x$.
   The color coding is in principle the same as in Figs.~\ref{fig:CompAllt0}
   and \ref{fig:CompAllt5},
  but there are only per mil level differences between the proxies for $k_0=10/\Delta x$.}
 \end{center}\end{figure}

  We also note that even in the massless case,
  all the proxies for position become practically identical for large
  momentum $\hbar k_0\gg\hbar/\Delta x$.
  This is illustrated for $k_0=10/\Delta x$ in Fig.~\ref{fig:CompAllLargeK}.
  The color coding is in principle the same as in Figs.~\ref{fig:CompAllt0}
  and \ref{fig:CompAllt5}.
  However, the differences between the four proxies are at the per mil level for $k_0=10/\Delta x$.

\begin{figure}[htb]\begin{center}
\scalebox{1}{\includegraphics{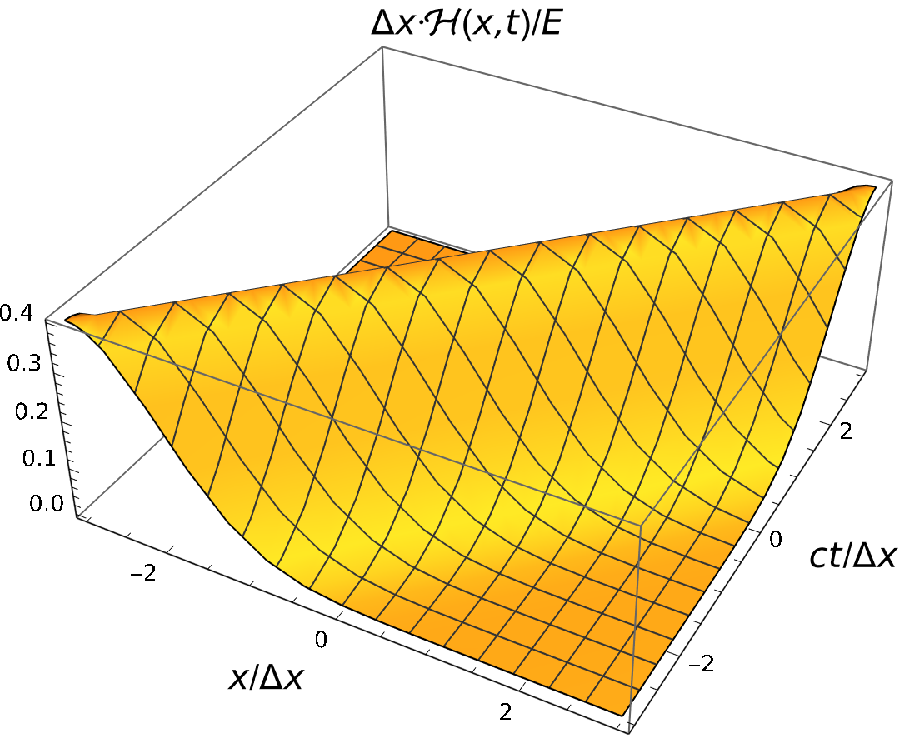}}
\caption{\label{fig:HLargeK}
  The normalized canonical energy density $\mathcal{H}(x,t)/E$
  of the massless Klein-Gordon field 
  with Gaussian wave packet (\ref{eq:Gausskk0}) and $k_0=10/\Delta x$.}
  \end{center}\end{figure}

All proxies for $k_0\Delta x>>1$ look like the normalized canonical energy
density $\mathcal{H}(x,t)/E$,
 see Fig.~\ref{fig:HLargeK} for $k_0=10/\Delta x$.
        
  We can explain this behavior from the asymptotic behavior of the wave packet
  $\psi(x,t)$ and the Klein--Gordon field $\phi(x,t)$ that follow from the wave packet
  (\ref{eq:Gausskk0}). The wave packet
  \begin{eqnarray} \nonumber
    \psi(x,t)&=&\left(\frac{\Delta x^2}{2\pi^3}\right)^{1/4}
    \int\!dk\,
    \exp[\mathrm{i}kx-\Delta x^2 (k-k_0)^2-\mathrm{i}c|k|t]
    \\ \nonumber
    &=&\frac{1}{2(2\pi\Delta x^2)^{1/4}}
    \exp\!\left(\mathrm{i}k_0(x-ct)-\frac{(x-ct)^2}{4\Delta x^2}\right)
 \\ \nonumber
 &&\times\left[1+\mathrm{erf}\!\left(\Delta x\cdot k_0+\mathrm{i}\frac{x-ct}{2\Delta x}\right)\right]
 \\ \nonumber
 &&+\frac{1}{2(2\pi\Delta x^2)^{1/4}}\exp\!\left(\mathrm{i}k_0(x+ct)-\frac{(x+ct)^2}{4\Delta x^2}\right)
 \\
 &&\times\left[1-\mathrm{erf}\!\left(\Delta x\cdot k_0+\mathrm{i}\frac{x+ct}{2\Delta x}\right)\right]
  \end{eqnarray}
  satisfies
  \begin{eqnarray} \label{eq:asympsi}
    \psi(x,t)&\to&\frac{1}{(2\pi\Delta x^2)^{1/4}}
    \exp\!\left(\mathrm{i}k_0(x-t)-\frac{(x-ct)^2}{4\Delta x^2}\right)
  \end{eqnarray}
  for $\Delta x\cdot k_0\gg 1$.
  The corresponding Klein--Gordon field (here displayed for $k_0\ge 0$)
    \begin{eqnarray} \nonumber
    \phi(x,t)&=&\left(\frac{\Delta x^2}{8\pi^3 c^2}\right)^{1/4}
    \int\!\frac{dk}{\sqrt{|k|}}\,
    \exp[\mathrm{i}kx-\Delta x^2 (k-k_0)^2-\mathrm{i}c|k|t]
    \\ \nonumber
    &=&\frac{1}{4}\left(\frac{\pi}{2c^2\Delta x^2}\right)^{1/4}
    \exp\!\left(-\,\frac{\Delta x^2 k_0^2}{2}\right)
     \exp\!\left(\mathrm{i}k_0\frac{x-ct}{2}-\frac{(x-ct)^2}{8\Delta x^2}\right)
     \\ \nonumber
     &&\times\sqrt{2\Delta x^2 k_0+\mathrm{i}(x-ct)}
     \left[I_{-\frac{1}{4}}\!\left(\frac{[2\Delta x^2k_0+\mathrm{i}(x-ct)]^2}{8\Delta x^2}\right)
       \right.
       \\ \nonumber
     &&+\left.I_{\frac{1}{4}}\!\left(\frac{[2\Delta x^2k_0+\mathrm{i}(x-ct)]^2}{8\Delta x^2}\right)
       \right]
 \\ \nonumber
 &&+\frac{1}{2(8\pi^3c^2\Delta x^2)^{1/4}}\exp\!\left(-\,\frac{\Delta x^2 k_0^2}{2}\right)
 \exp\!\left(\mathrm{i}k_0\frac{x+ct}{2}-\frac{(x+t)^2}{8\Delta x^2}\right)
     \\ \nonumber
     &&\times\sqrt{2\Delta x^2 k_0+\mathrm{i}(x+ct)}
     K_{\frac{1}{4}}\!\left(\frac{[2\Delta x^2k_0+\mathrm{i}(x+ct)]^2}{8\Delta x^2}\right)
  \end{eqnarray}
  satisfies
  \begin{eqnarray} \label{eq:asymphi}
    \phi(x,t)&\to&
    \frac{(\Delta x^2/2\pi c^2)^{1/4}}{\sqrt{2\Delta x^2 k_0+\mathrm{i}(x-ct)}}
    \exp\!\left(\mathrm{i}k_0(x-ct)-\frac{(x-ct)^2}{4\Delta x^2}\right)
  \end{eqnarray}
  for $\Delta x\cdot k_0\gg 1$.

  As a consequence of (\ref{eq:asympsi}) and (\ref{eq:asymphi}), all the proxies for position satisfy
  \begin{eqnarray}\nonumber
    \mathcal{H}(x,t)/E&\to&\varrho(x,t)/q\to\tilde{\mathcal{H}}(x,t)/E\to|\psi(x,t)|^2
    \\ \label{eq:largeKproxies}
    &\to&\frac{1}{\sqrt{2\pi\Delta x^2}}
    \exp\!\left(-\,\frac{(x-ct)^2}{2\Delta x^2}\right)
  \end{eqnarray}
  for $\Delta x\cdot k_0\gg 1$. In spite of the nonlocal relations between the position proxies,
  they do yield the same results even in the massless limit if $\Delta x\cdot k_0\gg 1$.
  We can understand this behavior also directly from Eq.~(\ref{eq:conck0}), which yields
  $E\simeq\hbar\omega(\bm{k}_0)$ and
  \begin{equation}\label{eq:genborn2}
    \mathcal{H}(\bm{x},t)/E\simeq\varrho(\bm{x},t)/q\simeq\tilde{\mathcal{H}}(\bm{x},t)/E
    \simeq|\psi(\bm{x},t)|^2.
  \end{equation}

\section{Applications to Dirac fields\label{sec:dirac}}

Our results for Klein--Gordon fields have direct impacts for Dirac fields, too.
To elucidate this, we recall that the free Dirac field can be written
in the form
\begin{eqnarray}\nonumber
\Psi(\bm{x},t)&=&\frac{1}{\sqrt{2\pi}^3}\int\!
d^3\bm{k}\sum_{s}
\big[b_s(\bm{k})u(\bm{k},s)\exp(\mathrm{i}k\cdot x)
\\ \label{eq:diracfs2}
&&+d_s^+(\bm{k})v(\bm{k},s)\exp(-\,\mathrm{i}k\cdot x)
\big]
\end{eqnarray}
with $k\cdot x\equiv\bm{k}\cdot\bm{x}-\omega(\bm{k})t$ and the basis
of normalized 4-spinors
\begin{eqnarray} \label{eq:u1ks}
u(\bm{k},{\scriptstyle\frac{1}{2}})&=&\frac{1}{\sqrt{2E(\bm{k})[E(\bm{k})+mc^2]}}
\left(\begin{array}{c}
E(\bm{k})+mc^2\\
0\\
\,\,\hbar ck_3\\
\,\,\,\hbar ck_+\\
\end{array}\right)\!,
\end{eqnarray}

\begin{eqnarray} \label{eq:u2ks}
  u(\bm{k},{\scriptstyle -\frac{1}{2}})&=&
  \frac{1}{\sqrt{2E(\bm{k})[E(\bm{k})+mc^2]}}
\left(\begin{array}{c}
0\\
E(\bm{k})+mc^2\\
\,\,\,\hbar ck_-\\
\!\!\!\!-\,\hbar ck_3\\
\end{array}\right)\!,
\end{eqnarray}

\begin{eqnarray} \label{eq:v2ks}
v(\bm{k},{\scriptstyle -\frac{1}{2}})
&=&\frac{1}{\sqrt{2E(\bm{k})[E(\bm{k})+mc^2]}}
\left(\begin{array}{c}
  \,\,\hbar ck_3\\
\,\,\,\hbar ck_+\\
E(\bm{k})+mc^2\\
0\\
\end{array}\right)\!,
\end{eqnarray}

\begin{eqnarray} \label{eq:v1ks}
  v(\bm{k},{\scriptstyle\frac{1}{2}})
  &=&\frac{1}{\sqrt{2E(\bm{k})[E(\bm{k})+mc^2]}}
 \left(\begin{array}{c}
    \,\,\,\hbar ck_-\\
\!\!\!\!-\,\hbar ck_3\\
0\\
E(\bm{k})+mc^2\\
\end{array}\right)\!.
\end{eqnarray}
Here $k_\pm=k_1\pm\mathrm{i}k_2$ was used.

In the Dirac case, the $\bm{x}$-space field for a single-particle state
\begin{eqnarray} \label{eq:1pdirac2}
\bm{|}\psi(t)\bm{\rangle}
&=&\sum_s\int\!d^3\bm{k}\,b_s^+(\bm{k})\bm{|}0\bm{\rangle}\psi_s(\bm{k})
\exp(-\,\mathrm{i}\omega_{\bm{k}}t),
\end{eqnarray}
\begin{equation}
\sum_s\int\!d^3\bm{k}\,|\psi_s(\bm{k})|^2=1,
  \end{equation}
has 4-spinor components
\begin{eqnarray} \nonumber
  \varphi_a(\bm{x},t)&=&\bm{\langle}0\bm{|}\Psi_a(\bm{x},t)\bm{|}\psi(0)\bm{\rangle}
  =\bm{\langle}0\bm{|}\Psi_a(\bm{x})\bm{|}\psi(t)\bm{\rangle}
  \\ \label{eq:phia}
  &=&\frac{1}{\sqrt{2\pi}^3}\int\!
  d^3\bm{k}\,\psi_s(\bm{k})u_a(\bm{k},s)
  \exp\!\left[\mathrm{i}(\bm{k}\cdot\bm{x}-\omega_{\bm{k}}t)\right]\!.
\end{eqnarray}
We assume a single-particle wave packet,
\begin{equation}
\psi_s(\bm{k})=\psi(\bm{k})\delta_{s,\frac{1}{2}}.
\end{equation}
The corresponding Fourier transformed wave packet
has components $\psi_s(\bm{x},t)=\psi(\bm{x},t)\delta_{s,\frac{1}{2}}$
with $\psi(\bm{x},t)$ given in Eq.~(\ref{eq:varphixt}).

We are focusing on the massless case and spin polarization along the
direction of momentum, i.e.~$k_\pm=0$.
The 4-spinor (\ref{eq:phia}) then becomes
\begin{equation}
  \varphi(\bm{x},t)=\psi(\bm{x},t)\frac{1}{\sqrt{2}}
  \left(\begin{array}{c}
1\\
0\\
1\\
0\\
\end{array}\right)\!.
  \end{equation}

In this case, the normalized charge density coincides with the
Born probability density from Fourier transform of the $\bm{k}$-space
wave packet,
\begin{equation}
  \varrho(\bm{x},t)/q=\varphi^+(\bm{x},t)\cdot\varphi(\bm{x},t)=|\psi(\bm{x},t)|^2,
\end{equation}
while the canonical energy density agrees with the
previously defined pseudo-density for energy after taking into account the Dirac equation,
\begin{eqnarray}\nonumber
  \mathcal{H}_{\varphi}&=&
  \frac{\hbar c}{2\mathrm{i}}\left(\varphi^+\gamma^0\bm{\gamma}\cdot\bm{\nabla}\varphi
    -\bm{\nabla}\varphi^+\cdot\gamma^0\bm{\gamma}\varphi\right)
    \\
    &=&\frac{\mathrm{i}\hbar}{2}\left(\psi^+\frac{\partial\psi}{\partial t}
    -\frac{\partial\psi^+}{\partial t}\psi\right)=\tilde{\mathcal{H}}.
\end{eqnarray}

These observations also apply in two spacetime dimensions.
A Dirac basis of $\gamma$ matrices is provided by
\begin{equation}
\gamma_0=\left(
\begin{array}{cc}
-\,{1} & {0}\\
\,\,\,\,\,{0} & {1}\\
\end{array}
\right)\!,\quad\gamma_1=\left(
\begin{array}{cc}
\,\,\,\,\,0 & 1\\
-\,1 & 0\\
\end{array}
\right)\!,
\end{equation}
and the general massless Dirac field is
\begin{eqnarray}\nonumber
\Psi({x},t)&=&\frac{1}{2\sqrt{\pi}}\int\!
dk
\left[b({k})
\left(
\begin{array}{c}
  1\\
  1\\
\end{array}
\right)
\exp[\mathrm{i}(kx-c|k|t)]
\right.
\\ 
&&+\!\left.d^+({k})
\left(
\begin{array}{cc}
\,\,\,\,\,1\\
-\,1\\
\end{array}
\right)
\exp[-\,\mathrm{i}(kx-c|k|t)]
\right]\!.
\end{eqnarray}

The Dirac spinor for the single-particle state
\begin{eqnarray} \label{eq:1pdirac2d}
\bm{|}\psi(t)\bm{\rangle}
&=&\int\!d\bm{k}\,b^+({k})\bm{|}0\bm{\rangle}\psi({k})
\exp(-\,\mathrm{i}c|k|t),
\end{eqnarray}
is
\begin{equation}
  \varphi({x},t)=\psi({x},t)\frac{1}{\sqrt{2}}
  \left(\begin{array}{c}
1\\
1\\
\end{array}\right)\!,
\end{equation}
and we find again
\begin{equation}
  \varrho({x},t)/q=|\psi({x},t)|^2,\quad
  \mathcal{H}_{\varphi}({x},t)=\tilde{\mathcal{H}}({x},t).
  \end{equation}

This means that the canonical
energy density of the massless fermion in
two spacetime dimensions with initial state
(\ref{eq:Gaussk0},\ref{eq:1pdirac2d}) is now displayed
in Figs.~\ref{fig:NormTildeHxt} and \ref{fig:NormTildeH0},
while the normalized charge density now agrees with the Born density
displayed in Figs.~\ref{fig:PsiKG0squared2} and \ref{fig:PsiKGsquared3}.

We note in particular that only the Born density is positive
  definite both for Klein--Gordon fields and for Dirac fields.

  \section{Conclusions\label{sec:conc}} 

  Even in the single-particle cases, normalized energy density remains the
  only positive definite position proxy for bosons while normalized charge
  density remains the only positive definite position proxy for fermions
  if we insist on expressions in terms of first-quantized fields.
  On the other hand, Born densities
  \begin{eqnarray} \label{eq:borndensities}
    &&\sum_s|\psi_s(\bm{x},t)|^2=\frac{1}{(2\pi)^3}
    \sum_s\left|\int\!d^3\bm{k}\,\psi_s(\bm{k},t)\exp(\mathrm{i}\bm{k}\cdot\bm{x})\right|^2
  \end{eqnarray}
  provide positive definite position proxies for any spin by construction,
  and we have seen
  that they remain tantalizingly close to the normalized energy and charge densities.
  The Born densities provide excellent approximations both to normalized energy
  density and normalized charge density in the 
  limits of low energy (\ref{eq:nonrelphi}) or strong localization in $\bm{k}$
  space (\ref{eq:conck0}).
  However, outside of these limits,
  the wave packets $\psi_s(\bm{x},t)$ are only indirectly related to
  the corresponding first-quantized fields, and therefore they do not satisfy
  the corresponding local evolution equations in background fields.

  We might consider the Born densities (\ref{eq:borndensities}) as a common
  underlying formalism for relativistic particle position. The indirect link
  of the Born densities to the local dynamics of the
  corresponding first-quantized fields necessitates position proxies which
  are directly related to the corresponding first-quantized fields.
  These position proxies are normalized energy densities for bosons
  or normalized charge densities for fermions.

  Alternatively, we could simply conclude that normalized energy densities
  are the correct position measures for bosons while normalized
  charge densities are the correct position measures for fermions,
  without assumption of a common underlying concept. Our findings
  cannot rule out this interpretation. However, the assumption of
  a principal difference in quantum mechanical formalism for bosons
  and fermions appears no less puzzling than the assumption of
  Born densities as a common concept that only indirectly links
  to local dynamics in relativistic regimes with large momentum
  uncertainty.

\subsection*{Acknowledgments}

We acknowledge support from the Natural Sciences and Engineering Research Council
of Canada.\\


\end{document}